\documentclass[12pt,letterpaper]{article}
\pdfoutput=1

\usepackage[pdftex]{graphicx}

\usepackage[utf8]{inputenc}

\usepackage{color}
\usepackage{slashed}
\usepackage{bbold,amsfonts,amssymb}
\usepackage{physics,amsfonts,amssymb,amsmath,amsrefs}

\usepackage{graphicx}
\usepackage{caption}
\usepackage{subcaption}

\usepackage{hyperref}

\numberwithin{equation}{section}

\newcommand{\be}{\begin{equation}}
\newcommand{\ee}{\end{equation}}
\newcommand{\ben}{\begin{displaymath}}
\newcommand{\een}{\end{displaymath}}
\newcommand{\bea}{\begin{eqnarray}}
\newcommand{\eea}{\end{eqnarray}}
\newcommand{\bean}{\begin{eqnarray*}}
\newcommand{\eean}{\end{eqnarray*}}

\def\e {\epsilon}

\newcommand{\calf}{\mbox{${\cal F}$}}

\newcommand{\calu}{\mbox{${\cal U}$}}
\newcommand{\calv}{\mbox{${\cal V}$}}

\renewcommand{\bra}[1]{\mbox{$\langle #1 |$}}
\renewcommand{\ket}[1]{\mbox{$| #1 \rangle$}}
\renewcommand{\braket}[2]{\mbox{$\langle #1  | #2 \rangle$}}

\newcommand{\ie}{{\it i.e.}}

\newcommand{\commentout}[1]{}

\usepackage{amsmath}

\newcommand{\beq}{\begin{equation}}
\newcommand{\eeq}{\end{equation}}
\newcommand{\beqr}{\begin{displaymath}}
\newcommand{\eeqr}{\end{displaymath}}
\newcommand{\beqa}{\begin{eqnarray}}
\newcommand{\eeqa}{\end{eqnarray}}
\newcommand{\beqar}{\begin{eqnarray*}}
\newcommand{\eeqar}{\end{eqnarray*}}
\renewcommand{\k}{\kappa}
\newcommand{\cH}{{\cal H}}

\newcommand{\cO}{{\cal O}}

\newcommand{\non}{\nonumber}

\newcommand{\cP}{{\cal P}}

\newcommand{\half}{\ensuremath{\frac{1}{2}}}

\begin{document}

\title{\Large \bf Circuit complexity near critical points}
\date{}

\author{
	Uday Sood$^1$,
	Martin Kruczenski$^{1,2}$\thanks{E-mail: \texttt{usood@purdue.edu, markru@purdue.edu.}} \\
	$^1$ Dep. of Physics and Astronomy,	and \\ $^2$ Purdue Quantum Science and Engineering Institute  \\
	 Purdue University, W. Lafayette, IN, USA. }

\maketitle

\begin{abstract}
	We consider the Bose-Hubbard model in two and three spatial dimensions and numerically compute the quantum circuit complexity of the ground state in the Mott insulator and superfluid phases using a mean field approximation with additional quadratic fluctuations. After mapping to a qubit system, the result is given by the complexity associated with a Bogoliubov transformation applied to the reference state taken to be the mean field ground state. In particular, the complexity has peaks at the $O(2)$ critical points where the system can be described by a relativistic quantum field theory. Given that we use a gaussian approximation, near criticality the numerical results agree with a free field theory calculation. To go beyond the gaussian approximation we use general scaling arguments that imply that, as we approach the critical point $t\rightarrow t_c$, there is a non-analytic behavior in the complexity $c_2(t)$ of the form $|c_2(t) - c_2(t_c)| \sim |t-t_c|^{\nu d}$, up to possible logarithmic corrections. Here $d$ is the number of spatial dimensions and $\nu$ is the usual critical exponent for the correlation length $\xi\sim|t-t_c|^{-\nu}$. As a check, for $d=2$ this agrees with the numerical computation if we use the gaussian critical exponent $\nu=\half$. Finally, using AdS/CFT methods, we study higher dimensional examples and confirm this scaling argument with non-gaussian exponent $\nu$ for strongly interacting theories that have a gravity dual.	  
\end{abstract}

\clearpage

\tableofcontents
\newpage

\section{Introduction}\label{sec1}
In quantum information theory, circuit complexity quantifies how hard it is to prepare a given state (state complexity) or a unitary (unitary complexity) by means of a given quantum circuit. The quantum circuit involves a sequence of elementary operations applied sequentially. A proper definition of complexity becomes possible after one chooses a universal set of gates that act as elementary operations, a reference state/unitary and a tolerance on the space of states/unitaries. Such a quantity is expected to capture interesting properties of the state that may not be seen in expectation values of local operators. In this paper we study the critical behavior of complexity in systems of interacting bosons. The focus is on the Superfluid-Mott Insulator transition, a quantum phase transition seen in the Bose Hubbard Model (BHM) which can be studied experimentally for example using cold atomic gases in optical lattice potentials.
The notion of complexity that we consider follows from identifying optimal circuits with geodesics on the space of circuits as was done in \cite{1,2,3} in the context of quantum computing and in \cite{4,5,6,7,8,9,10,11} in the context of quantum field theory. We note that other definitions based on the path integral have also been made for the complexity of a state in quantum field theory \cite{22,23,24}. These field theory developments were motivated by ideas of complexity in holography where complexity was defined via the gravitational dual of the field theory \cite{25,26,27,28,29,30,31}. 

In this paper, we study numerically the behavior of the complexity of the ground state across the phase diagram of the Bose-Hubbard model (BHM) on 2d and 3d cubic spatial lattices. We approximate the local bosonic Hilbert space at each site by a finite dimensional space of dimension $n$ and then map this finite-dimensional bosonic system to a space of qubits with constraints. For a similar construction, see \cite{32}. Our approach then uses the mean field approximation to compute the complexity. In that approximation, the Hamiltonian is written as a mean field Hamiltonian, describing non-interacting sites, plus fluctuations. The ground state is described as a mean field ground state with no entanglement between sites plus corrections that lead to correlations between sites. We define wave operators in terms of qubits and only keep corrections coming from second order terms in the Hamiltonian. The ground state is a condensate of waves which is a good approximation away from the critical points. 

To compute the complexity we first write both the exact and mean field Hamiltonians in terms of the same qubit (or spin $\half$) operators with $n$ internal indices which allows us to define the complexity of the exact ground state relative to the mean field ground state in the usual (quantum computing) manner. In a first approximation it is given by the complexity associated with the condensation of fluctuations and therefore can be computed by computing first the complexity associated with a Bogoliubov transformation. We call such a complexity $C_{QC}$. We then relate this to other complexity measures defined in the literature using discretized field theory models which we write with the superscripts $\kappa = 1,2,3...$ corresponding to a different functional choice that is made for different $\kappa$ in the definition of the complexity.
 
 We find that the behavior of complexity is different for the two different universality classes of quantum phase transitions in the BHM. Highest complexity is found at the O(2) critical points. Both of these facts follow from the fact that the complexity near phase transitions is especially sensitive to the low energy excitations. We highlight this fact by computing the contribution to the complexity from the low-lying modes of the system near the phase transitions. We also numerically find the scaling of the complexity as our model becomes critical in $d=2$ and $d=3$ spatial dimensions. Since we are using a gaussian approximation, the numerical critical exponents are the classical ones. 

The paper is organized as follows: in Section \ref{sec2}, we introduce the phases of the Bose Hubbard model and its mean field solution. In Section \ref{sec3}, we rewrite the Bose Hubbard Hamiltonian in terms of qubits allowing us to use the standard definitions of complexity. For the subsequent calculation we keep terms only quadratic in excitations above the mean field ground state. In Section \ref{sec4}, we compute the qubit complexity from the Bogoliubov transformation and relate this approach to a known free field theory approach in Section \ref{sec5}. In Section \ref{sec6}, we show how the low energy modes of the quadratic Hamiltonian reproduce the low energy spectrum of the BHM and discuss other features of our results for complexity. In Section \ref{sec7}, we discuss the scaling behavior of complexity that appears in holographic model using the CV conjecture for complexity \cite{25,26}. We give our conclusions in Section \ref{sec8} and discuss some ideas for future work.

\section{Bose Hubbard Model}\label{sec2}
Quantum Phase transitions occur when a coupling constant $g$ that measures the relative strength of two competing energy terms in the Hamiltonian is varied across some critical value $g=g_c$ where the ground state dependence on $g$ is non-analytic \cite{20},\cite{21},\cite{33}. Here the quantum phase transition we are interested in happens in a system of bosons in the background of a periodic potential and a two-particle repulsive interaction. The Hamiltonian is given by
\begin{align}
H = \int d^d x \psi^{\dagger}(x) ( -\frac{\hbar^2}{2m} \nabla^2 + \calv_0(x) + \calv_T(x))\psi(x) + \frac{\calu_0}{2} \int d^dx \psi^{\dagger}(x) \psi^{\dagger}(x) \psi(x) \psi(x)
\label{a1}
\end{align}
Here, $\calv_0$ is a periodic potential while $\calv_T$ is a slowly varying trapping potential that can be used to produce spatial inhomogeneities. The constant $\calu_0$ corresponds to a short-ranged repulsion between the bosons.
At low energies, we can keep only the lowest vibrational state at each minima of $\calv_0$ and the dynamics is given by the Bose Hubbard model \cite{18}:
\begin{align}
\cH = -J \sum_{\langle i,j \rangle} b_i^{\dagger} b_j + \sum_i \epsilon_i n_i + \frac{U}{2}\sum_{i} n_i (n_i - 1)
\label{bhmh}
\end{align}
Here $\langle i,j \rangle$ refers to all nearest neighbor pairs (if $i\neq j$ then both $\langle i,j \rangle$ and $\langle j,i \rangle$ appear in the sum), $J$ is a hopping term while $U$ is an on-site repulsion proportional to $\calu_0$. In this paper, we consider the homogeneous case and we set $\e_i = 0$ at each site. This system has a superfluid-Mott insulator transition first studied by \cite{19}. The insulating phase is characterized by zero compressibility, a gap in the excitation spectrum, and a quantized value of the density whereas the superfluid phase is gapless and the density varies continuously across the phase. Near the transition the correlation length is much larger than the lattice spacing and we can use a continuum field theory approximation. At generic transition points the relevant field theory is the Gross-Pitaevskii model but at particular points of the phase diagram the transition is in the universality class of the relativistic $O(2)$ quantum field theory in $(d+1)$-dimensions. The Bose Hubbard model can be realized in optical lattices and its properties like the nature of its excitation spectrum have been experimentally probed \cite{12,13,14,15,16}.

The phase diagram is well described by a mean-field approach \cite{20} where, as a first approximation, correlations are ignored and the ground state is taken as a product of the same state at each site:  $\ket{\Psi}=\bigotimes_{I}{\ket{\Phi}_I}$. Minimizing $\bra{\Psi}H \ket{\Psi}$ is equivalent to minimizing the functional 
\begin{align}
    \calf(\ket{\Phi}) = -(f J \bra{\Phi}b^{\dagger}\ket{\Phi}\bra{\Phi}b\ket{\Phi} - \bra{\Phi}\frac{U}{2}n(n-1) - \mu n\ket{\Phi}) -\lambda (\braket{\Phi}{\Phi} - 1)
\label{a2}
\end{align}
where we included the chemical potential $\mu$. Here $f$ is the number of nearest neighbors and $\lambda$ is a Lagrange multiplier. Minimizing this functional wrt $\bra{\Phi}$ reduces to finding the lowest energy eigenstate of the on-site mean field Hamiltonian 
\begin{align}
    \cH_{MF} = -f J (\phi b^{\dagger} + \phi^{*} b) + \frac{U}{2}n(n-1) - \mu n
\label{a3}
\end{align}
subject to the self-consistency condition $\phi = \bra{\Phi}b\ket{\Phi}$. This procedure gives the well-known lobes enclosing Mott insulator phases ($\phi=0$) in the phase diagram in the $(J/U, \mu/U)$ plane \cite{20}. The lobes are labeled by non-negative integers representing the density of bosons. The $O(2)$ critical points are at the tip of the lobes.   

\section{Mapping the Bose-Hubbard model to a system of qubits}\label{sec3}
One way to define complexity for any model is to map it to a system of qubits and then use a standard definition for the complexity, for example using the approach in \cite{2} \footnote{For a modified scheme of counting gates for the complexity using Suzuki-Trotter method, see \cite{49}}. The general properties of the complexity should remain the same as long as we choose gates involving only a few sites of the lattice. In particular, the peak in complexity that we see later at the critical point is due to the increased difficulty in creating long range correlations using few site operators. 
This definition requires an approximation where only a finite number of qubits are retained. Here we use a simple prescription valid for a finite lattice with $N$ sites where we keep a finite number $n$ of states in each site. If $\ket{\alpha}, \alpha=0\ldots n-1$ denote the possible states at each site, we map them to a system of $n$ qubits\footnote{We use the notation $\sigma^z \ket{0}=-\ket{0}$ and $\sigma^z \ket{1}=\ket{1}$} simply as
\begin{align}
&\ket{0} = \ket{10\ldots 0} \label{a4}\\
&\ket{1} = \ket{01\ldots 0} \label{a5}\\
& \ldots \nonumber \\
&\ket{n-1} = \ket{00\ldots 1} \label{a7}
\end{align}
Although not particularly efficient ($n$ qubits needed for $n$ states), it allows for a simple prescription to write any local operator as
\beq
\cO_i = \sum_{\alpha,\beta}  \bra\beta \cO_i \ket{\alpha}\, \sigma^+_{i,\beta}\,\sigma^-_{i,\alpha}
\label{a8}
\eeq
where $\sigma^\pm_{i,\alpha}$ are the usual Pauli matrices acting on a qubit $\alpha$ at site $i$. Physical states have occupation number one at each site $i$:
\beq
\sum_{\alpha} \sigma^+_{i,\alpha} \sigma^-_{i,\alpha} = \sum_\alpha \half(1+\sigma^z_{i,\alpha}) = 1, \ \ \ \ \forall i=1\ldots N
\label{quditConstraint}
\eeq
For the Bose-Hubbard model it is convenient to start with the mean-field Hamiltonian
\beq
\cH_{MF} = -  f J \sum_i (\phi b_i^\dagger + \phi^* b_i) + \frac{U}{2} \sum_i n_i(n_i-1) -\mu \sum_i n_i 
\label{a9}
\eeq
where $\phi=\bra{0} b\ket{0}$ where $\alpha=0$ denotes the ground state of $\cH_{MF}$  and the problem is solved self-consistently for $\phi$. The other values of $\alpha=1\ldots n-1$ denote the other eigenstates of the local mean field Hamiltonian with energies $\epsilon_\alpha$. By adding and subtracting the mean field Hamiltonian we can write the full Hamiltonian as
\begin{align}
\cH = \sum_{i,\alpha} \epsilon_\alpha \sigma^+_{i,\alpha}\,\sigma^-_{i,\alpha} + 
&f J \sum_{i, \alpha, \beta} (\phi B_{\alpha \beta} + \phi^* B^*_{\beta\alpha}) \, \sigma^+_{i,\alpha}\,\sigma^-_{i,\beta}
\label{bhqubith}
\\ -  & J \sum_{\substack{\langle ij \rangle \\ \alpha,\beta  \\ \alpha',\beta'}} B_{\alpha\beta}B^*_{\beta'\alpha'} \, \sigma^+_{i,\alpha}\,\sigma^-_{i,\beta}\, \sigma^+_{j,\alpha'}\,\sigma^-_{j,\beta'}
\nonumber
\end{align}
where $B_{\beta\alpha}=\bra{\beta} b^\dagger\ket{\alpha}$. Here $i$ is summed over all sites $i=1\ldots N$ and greek indices are summed over a basis of $\cH_{MF}$ eigenstates $\alpha=0\ldots n-1$. It is now convenient to split the sum over $\alpha$ into the contribution from the ground state $\alpha=0$ and the rest. The Hamiltonian can then be written as
\beq
\cH = \cH^{(0)}+\cH^{(1)}+\cH^{(2)}+\cH^{(3)}+\cH^{(4)}
\label{a10}
\eeq
where the index denotes the number of $\sigma_{i,\alpha\neq0}$ that each term contains. For example we have
\begin{align}
\cH^{(0)} &= N \epsilon_0 + fJN |\phi|^2 \label{a11}\\
\cH^{(1)} &= 0 \label{a12}\\
\cH^{(2)} &= \sum_{i,\alpha} (\epsilon_\alpha-\epsilon_0) \sigma^+_{i,\alpha}\,\sigma^-_{i,\alpha} \label{a13} \\
& -  J \sum_{\langle ij \rangle, \alpha,\beta} B_{\alpha0}B^*_{\beta0}  \, \sigma^+_{i,\alpha}\,\sigma^-_{i,0}\,\sigma^+_{j,0}\,\sigma^-_{j,\beta}
-  J \sum_{\langle ij \rangle, \alpha,\beta} B_{0\alpha}B^*_{0\beta}  \, \sigma^+_{i,0}\,\sigma^-_{i,\alpha}\,\sigma^+_{j,\beta}\,\sigma^-_{j,0}   
\nonumber
\\  
&- J \sum_{\langle ij \rangle, \alpha,\beta} B_{\alpha0}B^*_{0\beta}  \, \sigma^+_{i,\alpha}\,\sigma^-_{i,0}\,\sigma^+_{j,\beta}\,\sigma^-_{j,0}
-  J \sum_{\langle ij \rangle, \alpha,\beta} B_{0\alpha}B^*_{\beta0}  \, \sigma^+_{i,0}\,\sigma^-_{i,\alpha}\,\sigma^+_{j,0}\,\sigma^-_{j,\beta} \label{a14}\nonumber
\end{align}
From now on, sums over greek indices are taken from $1$ to $n-1$ (ground state omitted).
It is fairly easy to compute $\cH^{(3)}$ and $\cH^{(4)}$ also but we will not need their concrete expressions in this paper.
To continue we define the ``spin-wave'' operators 
\beq
\gamma_{k,\alpha} = \frac{1}{\sqrt{N}} \sum_j e^{-ijk} \sigma^+_{j,0}\,\sigma^-_{j,\alpha} 
\label{a15}
\eeq
that obey the commutation relations
\beq
[\gamma_{k,\alpha}, \gamma^\dagger_{k'\beta}] = \delta_{kk'}\delta_{\alpha\beta} -\frac{1}{N} \sum_j e^{-ij(k-k') }\sigma^+_{j,\beta}\,\sigma^-_{j,\alpha} -\frac{1}{N} \delta_{\alpha\beta} \sum_{j,\gamma} e^{-ij(k-k') }\sigma^+_{j,\gamma}\,\sigma^-_{j,\gamma} 
\label{a16}
\eeq
When $N$ is large ($N\gg 1$) and we work with states with only a few excitations we can drop the quadratic terms in the commutation relation. This is the usual approximation where we ignore scattering of spin waves. Then
\beq
[\gamma_{k,\alpha}, \gamma^\dagger_{k'\beta}] \simeq \delta_{kk'}\delta_{\alpha\beta} 
\label{bhcommutator}
\eeq
and these operators behave as unconstrained bosonic oscillators. Again, the initial qubits are constrained because we cannot excite more than one qubit at a given site but the spin waves are not since the probability of two spin waves exciting the same site is negligible for a large lattice (in a first approximation).
The quadratic Hamiltonian becomes
\begin{align}
\mathcal{H}^{(2)} &= \frac{1}{2} \sum_{k,\alpha \beta} \left( M_{\alpha \beta}(k) \gamma_{k,\alpha}^{\dagger} \gamma_{k,\alpha} + M^{*}_{\alpha \beta}(k)\gamma_{-k,\alpha} \gamma^{\dagger}_{-k,\beta} \right)  \nonumber  \nonumber\\
&+ \frac{1}{2} \sum_{k,\alpha \beta} \left( P_{\alpha \beta}(k) \gamma^{\dagger}_{k,\alpha}\gamma^{\dagger}_{-k,\beta} + P^{*}_{\alpha \beta}\gamma_{-k,\alpha} \gamma_{k,\beta}  \right)  \nonumber  \\
&= \frac{1}{2} \sum_{k,\alpha \beta} \begin{pmatrix} 
\gamma^{\dagger}_k & \gamma_{-k}
\end{pmatrix}
\begin{pmatrix}
M(k) & P(k) \\
P^{*}(k) & M^{*}(k)
\end{pmatrix}
\begin{pmatrix}
\gamma_{k}  \\
\gamma^{\dagger}_{-k}
\end{pmatrix}  \nonumber \\
&= \frac{1}{2} \sum_{k,\alpha \beta} \Gamma^{\dagger}_{k} \mathcal{H}(k) \Gamma_k
\label{quadraticbh}
\end{align}
where
\begin{align}
M_{\alpha \beta}(k)&=\left[(\epsilon_\alpha-\epsilon_0)\delta_{\alpha\beta} - Jf \eta_k (B_{\alpha0}B^*_{\beta0} + B_{0\beta}B^*_{0\alpha})\right]
\\
P_{\alpha\beta}(k)&= - J f \eta_k \left( B_{0\alpha} B^*_{\beta0} + B_{0 \beta} B^*_{\alpha0} \right) \label{a17}
\\
\eta_k &= \frac{1}{f}\sum_{a} e^{ika} \label{a18}
\end{align}
Here $a$ runs over all nearest neighbors. For a square lattice as we use here $\eta_k=\frac{1}{d}\sum_{j=1}^d \cos k_j$ is real. The $B$-matrices can also be taken to be real. Therefore, $\mathcal{H}(k)$ is real. Now the ground state for this ``spin wave'' system of $\gamma_k$ bosons can be computed by using a canonical transformation with a particular form. We need a transformation G such that $\Gamma_k = G(k) \Lambda_k$ with $G(k)$ of the form $G = \begin{pmatrix}
u & v \\
v & u
\end{pmatrix}
$ where $u$ and $v$ are $(n-1) \times (n-1)$ real matrices. For the transformation to be canonical i.e preserve the commutation relation, it has to preserve the metric $\kappa = \begin{pmatrix}
1 & 0 \\
0 & -1
\end{pmatrix}$ on this internal $2(n-1) \times 2(n-1)$ space.
\begin{align}
G \kappa G^{t} = \kappa
\label{a20}
\end{align}
where $G^{t}$ indicate the transpose of $G$. The set of matrices with the prescribed form obeying the above constraint form a group isomorphic to $GL(n-1, R)$. Starting with any element in $GL(n-1,R)$ $u_+$ and its inverse $u_+^{-1}$, we can construct a matrix $G(u_+)$ that satisfies the constraint \eqref{a20}. For example, take a matrix $u_{+} \in GL(n-1,R)$. Then the choice 
\begin{align}
u&= \frac{1}{2} (u_{+} + (u_{+}^{-1})^{t})  \label{a21} \\
v&=\frac{1}{2} (u_{+}  - (u_{+}^{-1})^{t}) \label{a22}
\end{align}
gives us a $G(u_{+})$ that preserves $\kappa$. On the other hand , a constraint-satisfying $G$ also uniquely specifies an element of $GL(n-1, R)$. The \eqref{a20} constraint is equivalent to 
\begin{align}
&uu^{t} - vv^{t} = 1 \label{a23} \\
&uv^{t} = vu^{t}   \label{a24}
\end{align}
which are the symmetric and anti-symmetric parts of the identity $(u + v)(u - v)^{t} = 1$. Therefore, the matrix $u_+ = u + v$ is an invertible real matrix. We use the singular value decomposition of matrices in $GL(n-1,R)$. For a choice of orthogonal matrices $\cal O$ and $\cal U$, we can write 
\begin{align}
u_{+} = \mathcal{O} e^{\theta} \mathcal{U}^{t}
\label{a25}
\end{align}
with a diagonal matrix $\theta$ since the singular values of $u_{+}$ are positive. Then $u$ and $v$ are of the form, 
\begin{align}
u&= \mathcal{O} \cosh \theta\, \mathcal{U}^{t}   \label{a26}\\
v&= \mathcal{O} \sinh \theta\, \mathcal{U}^{t}   \label{a27}
\end{align}
and $G$ is of the form
\begin{align}
G = \begin{pmatrix}
\mathcal{O} & 0 \\
0 & \mathcal{O} 
\end{pmatrix}
\begin{pmatrix}
\cosh \theta & \sinh \theta \\
\sinh \theta & \cosh \theta 
\end{pmatrix}
\begin{pmatrix}
\mathcal{U}^{t} & 0  \\
0 & \mathcal{U}^{t} 
\end{pmatrix}
\label{a33}
\end{align}
Plugging this transformation into the Hamiltonian, we obtain
\begin{align}
\mathcal{H}^{(2)} &=  \frac{1}{2} \sum_{k,\alpha \beta} \Lambda^{\dagger}_k \kappa G_k^{-1} \kappa
\mathcal{H}_k
G_k \Lambda_k   
\label{a34}
\end{align}
So we see that G has to be chosen s.t. it diagonalises $\kappa \mathcal{H}$ through a similarity transform $G^{-1} \kappa \mathcal{H} G = \mathcal{H}_{d}$. The eigenvalue problem for $\kappa \mathcal{H}$ has eigenvalues that come in pairs and $\mathcal{H}_{d}$ is of the form $\begin{pmatrix}
\omega & 0  \\
0 & -\omega
\end{pmatrix}$ where $\omega$ is an $(n-1) \times (n-1)$ diagonal matrix with positive entries $\omega_\alpha$. This gives an $\mathcal{H}^{(2)}$ that is diagonal in the $\lambda_k$ oscillators so that the target state satisfies $\lambda_{k, \alpha} \ket{\Omega} = 0$  $\forall k, \alpha$. This amounts to 
\begin{align}
(\mathcal{U}_k \cosh \theta_{k} \mathcal{O}_{k}^{t} \gamma_k - \mathcal{U}_{k} \sinh \theta_k \mathcal{O}_{k}^{t} \gamma^{\dagger}_{-k})_{\alpha} \ket{\Omega} = 0
\label{a35}
\end{align}
Left multiplying by $\mathcal{U}_{k}^{t}$, we get 
\begin{align}
(\cosh \theta_{k,\alpha} \tilde{\gamma}_{k, \alpha} - \sinh \theta_{k, \alpha} \tilde{\gamma}^{\dagger}_{-k, \alpha} ) \ket{\Omega} = 0
\label{a36}
\end{align}
where we define $ \tilde{\gamma}_{k, \alpha}$ operators as orthogonally related to $\gamma_{k, \alpha}$, $\tilde{\gamma}_{k, \alpha} = (\mathcal{O}^{t})_{\alpha \beta} \gamma_{k, \beta}$. Once we fix the norm of $\ket{\Omega}$ so that $\langle \Omega | \Omega \rangle = 1$, we have 
\begin{align}
\ket{\Omega} &= e^{\sum_{k, \alpha} \theta_{k, \alpha}(\tilde{\gamma}_{k, \alpha}^{\dagger}  \tilde{\gamma}_{-k, \alpha}^{\dagger} - \tilde{\gamma}_{k, \alpha}\tilde{\gamma}_{-k, \alpha})} \ket{0}  \\
&= e^{\sum_{k, \beta \delta} \Theta_{k, \beta \delta} (\gamma^{\dagger}_{k, \beta} \gamma^{\dagger}_{-\k, \delta} - \gamma_{k, \beta} \gamma_{-k, \delta})} \ket{0}
\label{bogoliubovt}
\end{align}
Here, the sum over $k$ only involves half of the total momentum modes and the matrix $\Theta = \mathcal{O}\theta \mathcal{O}^{t}$ where we used the $k \rightarrow -k$ symmetry.

Thus, the corrected ground state differs from the mean field ground state by the condensation of certain $k$ modes. 
\beq
\cH^{(2)} = \sum_{k,\alpha} \omega_{k,\alpha} \lambda_{k,\alpha}^\dagger \lambda_{k,\alpha}
\label{a38}
\eeq
and the ground state $\ket{\Omega}$ of $\cH^{(2)}$ is obtained by solving
\beq
\lambda_{k,\alpha} \ket{\Omega} =0
\label{a39}
\eeq
In the following section we compute the complexity of $\ket{\Omega}$ relative to $\ket{0}$ which, in this approximation, is just the complexity associated with such a condensation. 

\section{Computation of circuit complexity}\label{sec4}

 Given a reference state $\ket{0}$ we are interested in defining and then computing the complexity of another state $\ket{\Omega}$. According to the procedure described in \cite{2}, first we find a parameter dependent unitary $U(\tau)$ such that 
\beqa
  U(0) &=& \mathbb{1} \\
  U(1) \ket{0} &=& \ket{\Omega}
  \label{Ueq}
\eeqa
This unitary can also be written as
\beq
 U(\tau) = \hat{\cP} \left\{e^{-i \int_0^{\tau} H(\tau') d\tau'}  \right\}
\label{a40}
\eeq
where we defined a hermitian ``Complexity Hamiltonian'' or control function
\beq
 H(\tau) = i \partial_\tau U U^\dagger
\label{a41}
\eeq
It should be noted that this Hamiltonian has no relation to the actual Hamiltonian that determines the dynamics of the system. It is used only to compute the complexity of the ground state given the reference state. For a system of $M$ qubits the most general such Hamiltonian is 
\beq
 H(\tau) = \sum_{\{\sigma\}} h_{\{\sigma\}}(\tau)\ V_{\{\sigma\}}
\label{a42}
\eeq
where $\{\sigma\}$ represents a string of Pauli matrices acting on different qubits (including the identity there are $4^M$ such operators). Then we define a weight associated with such unitary path:
\beq
  d(U) = \int_0^1 \sqrt{ \sum_{\{\sigma\}}  p_{\{\sigma\}}^{2} \left| h_{\{\sigma\}}(\tau) \right|^2\, }\ d\tau
\label{a43}
\eeq 
where $p_{\{\sigma\}}$ is a, for the moment arbitrary, penalty assigned to each operator. Then, following \cite{2}, the complexity is defined as 
\beq
 C_{QC} = \mbox{min}_{U(\tau)}\ d(U)
\label{C2def}
\eeq
where we minimized over all unitaries\footnote{For a given $U(1)$ this quantity is the so called unitary complexity of the unitary U(1). The relative state complexity then has an additional step of minimizing over all unitaries $U(1)$ that connect $\ket{0}$ and $\ket{\Omega}$.} $U(\tau)$ that satisfy (\ref{Ueq}). The purpose of the $p_{\{\sigma\}}$ is to constrain the movement inside the unitary group only in certain``easy'' directions with small $p_{\{\sigma\}}$ based on the available elementary gates. This captures the notion of using elementary operations to construct unitaries and states and hence of gate complexity. Before continuing let us make some brief comments on the choice of penalties $p_{\{\sigma\}}$. First, if all $p_{\{\sigma\}}=1$ then we are dealing with the usual distance in $SU(2^M)$ and the complexity is the $SU(2^M)$ geodesic distance. In that case $H$ is independent of $\tau$ and given by an element of the Lie algebra such that $U=e^{-iH}$. Second, if we take two sets of weights $p^{(1)}_{\{\sigma\}}\ge p^{(2)}_{\{\sigma\}}$ then $C_{QC}^{(1)}\ge C_{QC}^{(2)}$, in particular, if all $p_{\{\sigma\}}\ge 1$ then the associated complexity is always larger than the geodesic distance.   

In the case of the lattice system studied here, once the space of states at each site is truncated to dimension $n$ it can be considered as a system of qudits with dimension n for which one and two qudits gates are universal \cite{48}. Notice also that, in the qubit picture we use, the most general one and two qudit operators can be written as
\beq
H(\tau) = \frac{1}{\sqrt{2}} \sum_{IJ,\alpha\beta,\gamma,\delta}  \Theta_{IJ,\alpha\beta\gamma\delta}(\tau)\ \sigma_{I,\alpha}^+ \sigma_{I,\beta}^- \sigma_{J,\gamma}^+ \sigma_{J,\delta}^- +\mbox{h.c.}
\label{twoqudit}
\eeq
where $I,J$ label the qudits (or sites) and this includes single qudit operators in view of the constraint (\ref{quditConstraint}), and any two qudit operator is given by an arbitrary $n^2\times n^2$ matrix. For that reason it is natural to choose the weights $p_{\{\sigma\}}$ to be one for such operators and infinite for all others. This is similar to keeping one and two qubit operators in a standard qubit system. 
With this choice of $p_{\{\sigma\}}$, the weight for such a Hamiltonian is then
\beq
d(\Theta) = \int_0^1 \sqrt{  \sum_{IJ,\alpha\beta,\gamma,\delta}  \left| \Theta_{IJ,\alpha\beta\gamma\delta}(\tau) \right|^2\, }\ d\tau
\label{a44}
\eeq
If $H(\tau)$ is independent of $\tau$ we get simply
\beq
  d^2(\Theta) = \sum_{IJ,\alpha,\beta,\gamma,\delta} \left| \Theta_{IJ,\alpha\beta\gamma\delta}\right|^2 
\label{a45}
\eeq
In principle we are interested in the case where the reference state $\ket{0}$ is the ground state of the mean field Hamiltonian and the state $\ket{\Omega}$ is the exact ground state. In fact, the qubit construction we made, gives a precise but difficult to evaluate definition of the complexity for any state. However, we are going to make the approximation of the previous section where we just consider the ground state of $\cH^{(2)}$ which is related to $\ket{0}$ by the condensation of certain momentum modes or equivalently by the Bogoliubov transformation given in (\ref{bogoliubovt}). Therefore there is a unitary of the form $U=e^{-iH}$ such that $\ket{\Omega}=U\ket{0}$ with
\beq
 H = \sum_{k,\beta\delta} \theta_{k,\beta\delta} (\gamma_{k,\beta}^\dagger\gamma_{-k,\delta}^\dagger -\gamma_{k,\beta}\gamma_{-k,\delta}) 
\label{a46}
\eeq
 namely of the form (\ref{twoqudit}) with
\beq
 \Theta_{IJ,\alpha 0 \beta 0} = \frac{i}{N} \sum_k \theta_{k,\alpha\beta} e^{i(I-J)k} 
\label{a47}
\eeq  
where we used $\theta^*_{k,\alpha\beta} = \theta_{-k,\beta\alpha}$. The complexity is therefore
\beqa
 C_{QC}^2 &=& \sum_{IJ,\alpha,\beta} \left| \Theta_{IJ,\alpha 0 \beta 0}\right|^2  \\
       &=& \frac{1}{N} \sum _{IJ,\alpha\beta,kk'} \theta_{k,\alpha\beta}\, \theta^*_{k',\alpha\beta} \, e^{i(I-J)(k-k')} \\
       	&=& \sum_{\alpha\beta,k} |\theta_{k,\alpha\beta}|^2 = \sum_{k,\alpha} |\theta_{k,\alpha}|^2  
\label{a48}
\eeqa 
Thus we find the formula that we use in this paper for the computation of the complexity in the approximation that the two states are related by the Bogoliubov transformation in (\ref{bogoliubovt}):
\beq
 C_{QC} = \sqrt{\sum_{k,\alpha} |\theta_{k,\alpha}|^2}  
\label{a49}
\eeq
Technically, this is an upper bound on the complexity since we have not proven mathematically that this gives the minimum but it is hard to see how deviating from this direct path from $\ket{0}$ to $\ket{\Omega}$ would a give smaller distance. Following \cite{4} we can define a slightly more generic complexity\footnote{Initially, complexities in \cite{4} had a constraint of positive homogeneity in the cost functions that required $C_{\kappa}$ to be of the form $(\sum_{\rm{p}} | \theta_p |^{\kappa})^{1/\kappa}$. However, this constraint meant that $C_{\kappa}$ did not scale like the spatial volume of the lattice and therefore Section 4 of \cite{4} considered complexities of the form \ref{a50}.}
\beq
 C_\kappa = \sum_{k,\alpha} |\theta_{k,\alpha}|^\kappa
\label{d1}
\eeq
 Notice however that only in the particular case $\kappa=2$ the steps in \eqref{a48} are valid namely $C_2=C_{QC}^2$. The complexities $C_{\kappa\ne2}$ are introduced for comparison with \cite{4} but their meaning in terms of the generic complexity Hamiltonian \eqref{twoqudit} is less clear.  

\section{Circuit Complexity associated with a Bogoliubov transformation}\label{sec5}

Given that the Hamiltonian \eqref{quadraticbh} is quadratic, it is natural to use a free field theory formalism to compute circuit complexity \cite{4} and compare it with the results from the qubit approach of the previous section. The free field theory approach in \cite{4} adapted the idea of geometrising the space of circuits to compute the $\kappa$-complexity for the ground state of free bosonic quantum field theories which is Gaussian in the field variables. The main result is the expression for the $\kappa$-complexity
\begin{align}
&C_{\kappa} = \frac{1}{2^{\kappa}} \sum_{\rm{p}} | \log (\tilde{\omega}_{\rm{p}} / \omega_0) |^{\kappa}  =   \sum_{\rm{p}} | \eta_p |^{\kappa} 
\label{a50}
\end{align}
where the sum is over all allowed momenta on the lattice, the frequencies $\tilde{\omega}_p$ are the normal modes of the lattice system and the second equality is just a definition of the $\eta_p$ parameters. The parameter $\kappa$ labels a set of choices for the cost function used in defining the complexity.
This result can also be interpreted as the relative complexity between the vacua of two sets of oscillators related by a Bogoliubov transformation. Let us illustrate this for a pair of oscillators. Consider the oscillators,
\begin{align}
&x_{j} = \sqrt{\frac{1}{2\omega_0}} (a_{j} + a_{j}^{\dagger}) \label{c51}\\
&p_{j} = i \sqrt{\frac{\omega_0}{2}} (a_{j}^{\dagger} - a_{j}) 
\label{a51}
\end{align}
for $j=1,2$ and their counterparts which are labelled by a ``momentum'' $k=+,-$.
\begin{align}
&x_{k} = \sqrt{\frac{1}{2\omega_{k}}} (a_{k} + a_{k}^{\dagger}) \label{c52}\\
&p_{k} = i \sqrt{\frac{\omega_{k}}{2}} (a_{k}^{\dagger} - a_{k})
\label{a52}
\end{align}
The annihilation operators defined this way are designed to annihilate two different states.
\begin{align}
&a_{j}(x_j,\partial_j) e^{-\frac{\omega_0}{2}(x_1^2 + x_2 ^2)}=0 \label{c53}\\
&a_{k}(x_k, \partial_k) e^{-\frac{1}{2}(\omega_+ x_+^2 +\omega_- x_- ^2)}=0
\label{a53}
\end{align}
The state in \eqref{c53} is a reference state $\ket{0}$ with frequency $\omega_0$ and the state in \eqref{a53} is a target state $\ket{\Omega}$ with normal frequencies $\omega_{\pm}$.
If the two pairs of oscillators are directly related by a Bogoliubov transformation, we can read off the complexity from the transformation itself using \eqref{a49}. Corresponding to $a_j$ we can define the linear combinations $A_k$ with $A_{\pm} = \frac{a_1 \pm a_2}{2}$ and $A_{\pm} \ket{0} = 0$.
In terms of the ``momentum space field variables'', a relation of the sort 
\begin{align}
A_{k}=\sqrt{\frac{\omega_0}{2}}(x_{k} + \frac{i}{\omega_0} p_{k})
\label{a54}
\end{align}
follows. A Bogoliubov transformation to normal mode oscillators has the form
\begin{align}
&a_+ = \cosh{\theta_+} A_+ + \sinh{\theta_+}A_{+}^{\dagger} \label{b55} \\
&a_- = \cosh{\theta_-} A_- + \sinh{\theta_-q}A_-^{\dagger}   \label{a55}
\end{align}
Then for this case,
\begin{align}
&\theta_+ = \eta_+  \label{b56}\\
&\theta_- = \eta_-  \label{a56}
\end{align}
This can be seen by noting that 
\begin{align}
&a_{+}=\sqrt{\frac{\omega_+}{2}} (x_+ + \frac{i}{\omega_{+}}p_+ )  \label{b57}\\
&a_{+} = \frac{1}{2}( (\sqrt{\frac{\omega_+}{\omega_0}} + \sqrt{\frac{\omega_0}{\omega_+}})A_+ +  (\sqrt{\frac{\omega_+}{\omega_0}} - \sqrt{\frac{\omega_0}{\omega_+}})A_+ ^{\dagger})  \label{a57}
\end{align}
Comparing with $a_+ = \cosh{\theta_+} A_+ + \sinh{\theta_+}A_+^{\dagger}$, we see that $e^{\theta_+} = \sqrt{\frac{\omega_+}{\omega_0}}$ and hence $\theta_+ = \frac{1}{2} \log (\frac{\omega_+}{\omega_0}) = \eta_+$. Similarly, for $\theta_-$ one finds that $\theta_- =\frac{1}{2} \log (\frac{\omega_-}{\omega_0})= \eta_-$. This shows that in the free field theory calculation of \cite{4} the target state and the reference state are related by a Bogoliubov transformation and therefore we can apply \eqref{d1}. It gives the same answer for the complexity as the one found in \cite{4} providing a further check of our results.  
   \\
Further, this allows us to compute the complexity of the ground state for any system with the schematic Hamiltonian $h_0$ with an energy scale $\epsilon_0$
\begin{align}
h_{0}/\epsilon_0 =  A_{+} ^{\dagger} A_+ + A_-^{\dagger} A_- + \lambda (A_+ ^{\dagger} A_-^{\dagger} + A_- A_+) 
\label{a59}
\end{align}
We make the inverse of the Bogoliubov transformations in \eqref{b55}, \eqref{a55} with $\theta_+ = \theta_- = \frac{1}{2} \tanh^{-1}{\lambda}$ to get 
\begin{align}
h_{0}/\epsilon_0 =  \sqrt{1-\lambda^2} - 1 + \sqrt{1-\lambda^2} (a_+^{\dagger} a_+ + a_- ^{\dagger} a_-) 
\label{a60}
\end{align}
Then, the unitary is generated by an operator of the form $(A_{+}^{\dagger}A_{-}^{\dagger} - A_{+}A_{-})$ and the $\kappa$-complexity $C_{\kappa}$ for this 2-mode model is $\frac{1}{2^{\kappa -1}} |\tanh^{-1}{\lambda}|^{\kappa}$. \\
As a simple example of this, we can apply the above idea to a weakly interacting homogeneous bose gas. We consider the Hamiltonian \cite{17} 
\begin{align}
H = \sum_{p} \frac{p^2}{2m} a^{\dagger}_p a_p + \frac{U_0}{2V} \sum_{p,q,r} a^{\dagger}_{p+r} a^{\dagger} _{q-r} a_p a_q
\label{a61}
\end{align}
that captures the physics of interacting bosons in the limit of small momentum. Then, consider $N_0$, $N$, $V$ $\to$ $\infty$ with the densities held fixed at $n_0 \neq 0$ and $n$. Here the subscript $0$ represents the $\mathbf{p=0}$ mode. Expanding in $a_p, a^{\dagger}_q$ with non-zero momentum (or equivalently in fluctuations $\delta \phi$ about the condensed state) up to quadratic order, we end up with the Hamiltonian 
\begin{align}
H = E^{(0)} + \sum_{\bold{p}>0}  [(\frac{p^2}{2m} + U)(a^{\dagger}_p a_p +a^{\dagger}_{-p} a_{-p}) + U(a^{\dagger}_p a^{\dagger}_{-p} + a_p a_{-p}) ]
\label{quadratic}
\end{align}
with $U = nU_0$. For every momentum $\bold{p}$, we have to perform a Bogoliubov transformation of the form \eqref{b55}, \eqref{a55}. We get a momentum dependent parameter $\theta_p$  
\begin{align}
\theta_p = \frac{1}{2} \tanh^{-1}\left( {\frac{2mU}{p^2 + 2mU}} \right)
\label{a62}
\end{align}
The total complexity for the ground state is then
\begin{align}
C_{\kappa} =  \sum_{p} |\theta_p|^{\kappa} \approx   V \int \frac{d^dp}{(2\pi)^d} |\theta(p)|^{\kappa}
\label{a63}
\end{align}
For the 3-dimensional gas, the integral for the density of $C_2$ which we denote as $c_2$ gives 
\begin{align}
c_2^{(d=3)}= \frac{1}{48\pi}(2-\log 4)(2mU)^{3/2}
\label{a64}
\end{align}
We see that complexity for a mode grows as we look at modes with small momentum. This is because the number of scaling gates (in the normal mode basis) increases with decreasing k in the circuit model of \cite{4}.
However, it should be noted that this approach can not directly be applied to study the behaviour of the complexity near the quantum phase transitions in the Bose Hubbard model because of the failure of the approximation (\ref{quadratic}) to capture the phase transition. This is where the results of Section \ref{sec3} become useful to us.

\section{Complexity in the Bose-Hubbard Model}\label{sec6}
\subsection{Physical Spectrum}
\begin{figure*}[h]
\centering
\begin{subfigure}[t]{0.4\textwidth}
\raggedleft
\includegraphics[width=\textwidth,height=4cm]{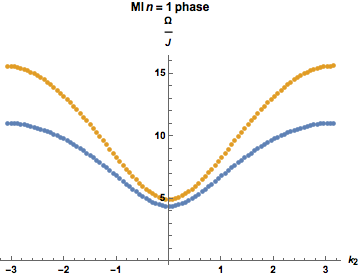}
\caption{$t=0.15$, $\overline{\mu}=\sqrt{2}-1$}
\end{subfigure}
\hfill
\begin{subfigure}[t]{0.4\textwidth}
\raggedright
\includegraphics[width=\textwidth,height=4cm]{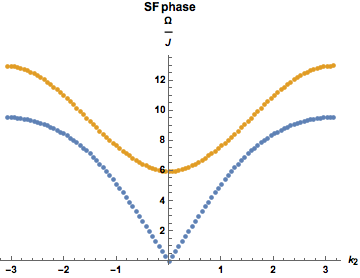}
\caption{$t=0.20$, $\overline{\mu}=\sqrt{2}-1$}
\end{subfigure}
\\
\begin{subfigure}[t]{0.4\textwidth}
\raggedleft
\includegraphics[width=\textwidth,height=4cm]{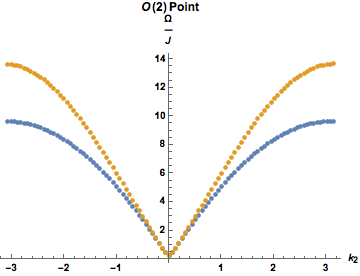}
\caption{$t=3-2 \sqrt{2}$, $\overline{\mu}=\sqrt{2}-1$}
\end{subfigure}
\hfill
\begin{subfigure}[t]{0.4\textwidth}
\raggedright
\includegraphics[width=\textwidth,height=4cm]{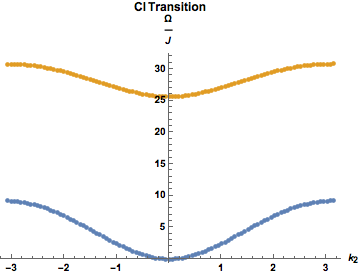}
\caption{$t=0.10$, $\overline{\mu}=0.77$}
\end{subfigure}

\caption{For $d=2$ spatial lattice, the lowest branches of the spectrum with $\Omega/J$ denoting the ratio of the eigenvalue of the Hamiltonian with the hopping amplitude. This quantity is plotted as a function of $k_2$ with $k_1=0$. The lattice is $100 \times 100$. The local Hilbert space has dimension 6.} 
\label{fig1}
\end{figure*}

We illustrate in this section how the excitations in the Hamiltonian (\ref{quadraticbh}) with the approximation of (\ref{bhcommutator}) reproduce the expected spectrum of the BH model, including its behaviour near the phase transitions. In \cite{19} the value of the dynamical exponent in this model was found to be two for the MI-SF transition at generic points, except near the O(2) point (tip of the lobe) where it is one highlighting the emergent Lorentz invariance at that point. Quantum critical points described by relativistic field theories like the O(2) point have been the subject of numerous studies in condensed matter physics \cite{35}-\cite{37}. For $d=2$ the coordination number is $f=4$. We define new variables $t=J f/U$ and $\overline{\mu}=\mu/U$. For the $n=1$ Mott insulating lobe, the perturbative value of the tip of the lobe is at $(t_c,\overline{\mu}_c) = (3-\sqrt{8},\sqrt{2}-1)$ \cite{34}.
We find an agreement with that and the self-consistency method of Section \ref{sec2}. We calculate the spectrum obtained from \eqref{quadraticbh}. This is presented in Figure \ref{fig1}. In particular 
Figure \ref{fig1}(a) displays the gapped modes characteristic of the Mott Insulator phase while Figure \ref{fig1}(b) in the SF phase has one gapless mode with a linear dispersion relation corresponding to the fact that it is a symmetry broken phase. Figure \ref{fig1}(c) and Figure \ref{fig1}(d) capture the spectrum at the two different kinds of phase transitions. Figure \ref{fig1}(c) has two linearly dispersing modes with the gapped amplitude mode in the SF phase also becoming gapless at this point. Figure \ref{fig1}(d) on the other hand has just one gapless mode which behaves differently and is not linear. In fact, it is easy to verify that for this mode, $\omega \sim k^2$ for small $k$ as expected.  
\begin{figure*}[h]
  \centering
  \begin{subfigure}[t]{0.4\textwidth}
  \raggedleft
  \includegraphics[width=\textwidth,height=5cm]{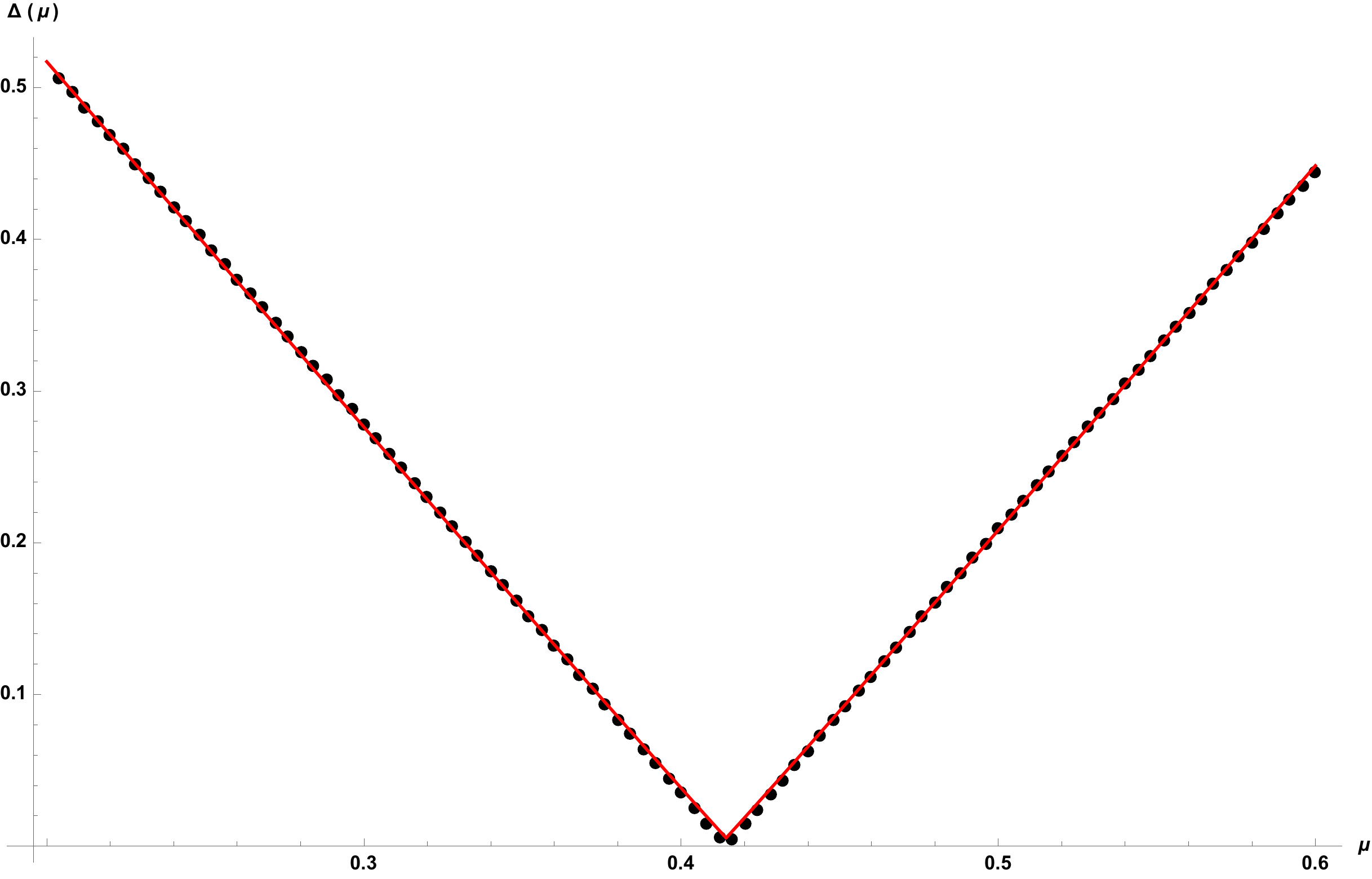}
  \end{subfigure}
  \hfill
  \begin{subfigure}[t]{0.4\textwidth}
  \raggedright
  \includegraphics[width=\textwidth,height=5cm]{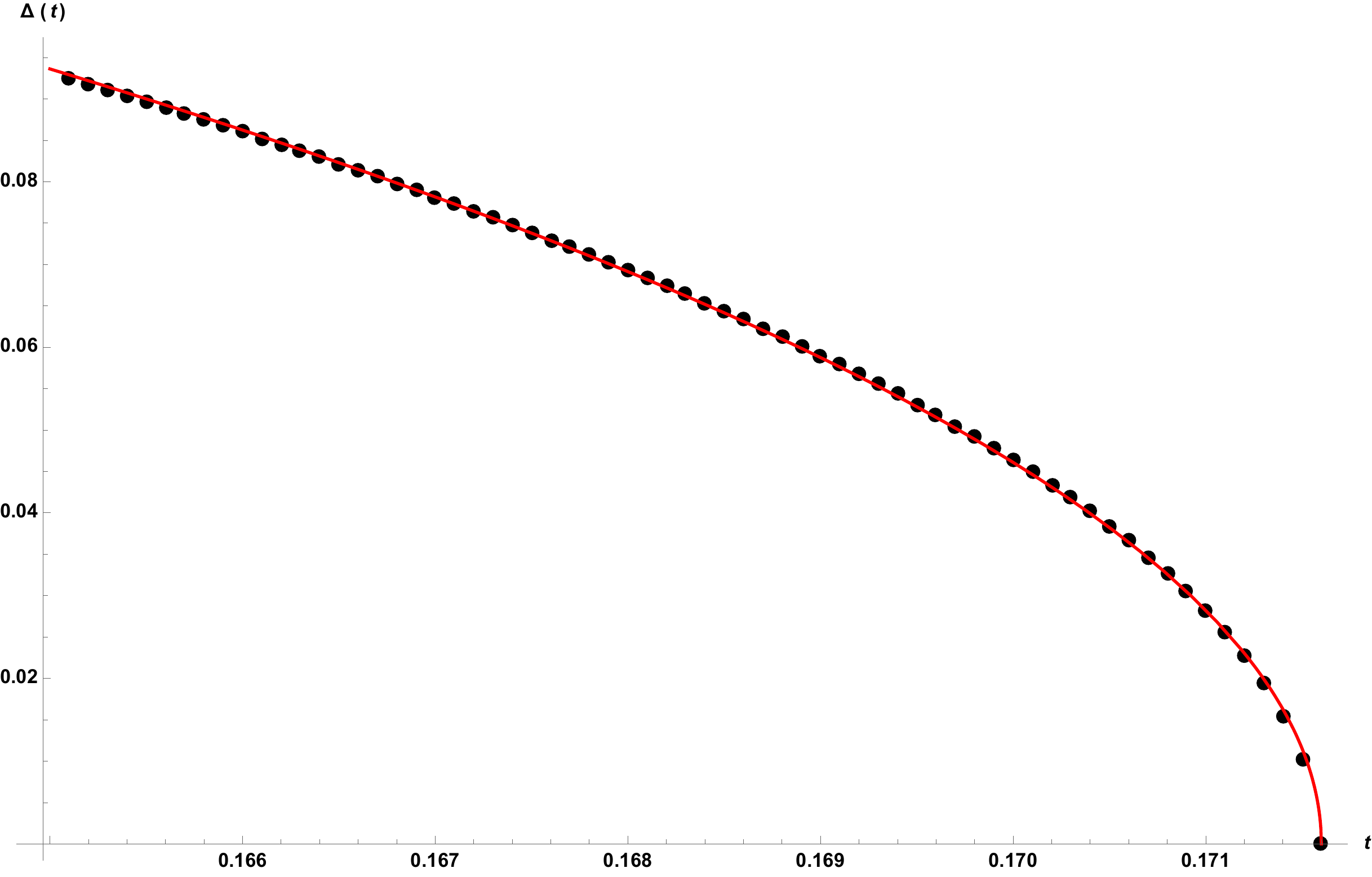}
  \end{subfigure}
  \caption{The scaling of the gap in the two different approaches to the $O(2)$ critical point gives an idea of the kind of exponents in the quadratic approximation. The left graph shows the behavior of the massive superfluid mode as we cross $\overline{\mu}_c$ with $t=t_c$. The red line is a curve linear in $\abs{\overline{\mu}-\overline{\mu}_c}$. The right graph is for $\overline{\mu}=\overline{\mu}_c$ and variation of $t$ in the mott insulator phase. The gap closes at $t_c$. The red curve is a fit with $(t_c-t)^{1/2}$ up to an overall coefficient.}
  \label{fig3}
\end{figure*}
 
\subsection{Complexity}
\begin{figure*}[h]
  \centering
  \begin{subfigure}[t]{0.4\textwidth}
  \raggedleft
  \includegraphics[width=\textwidth,height=5cm]{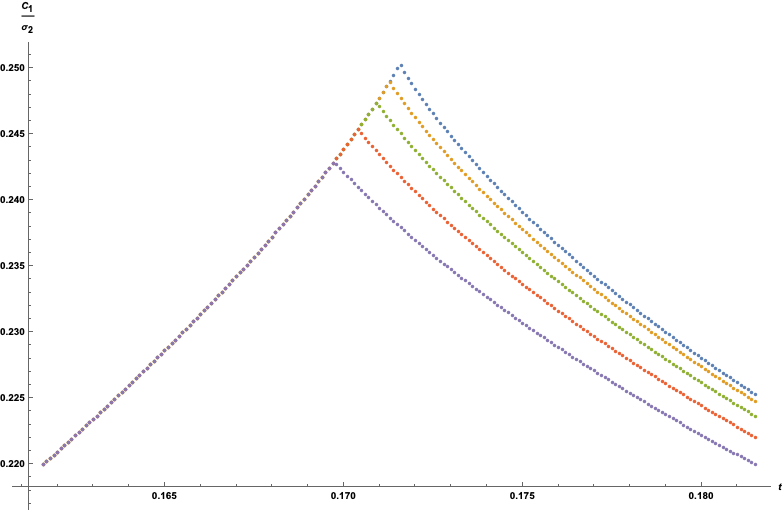}
  \end{subfigure}
  \hfill
  \begin{subfigure}[t]{0.4\textwidth}
  \raggedright
  \includegraphics[width=\textwidth,height=5cm]{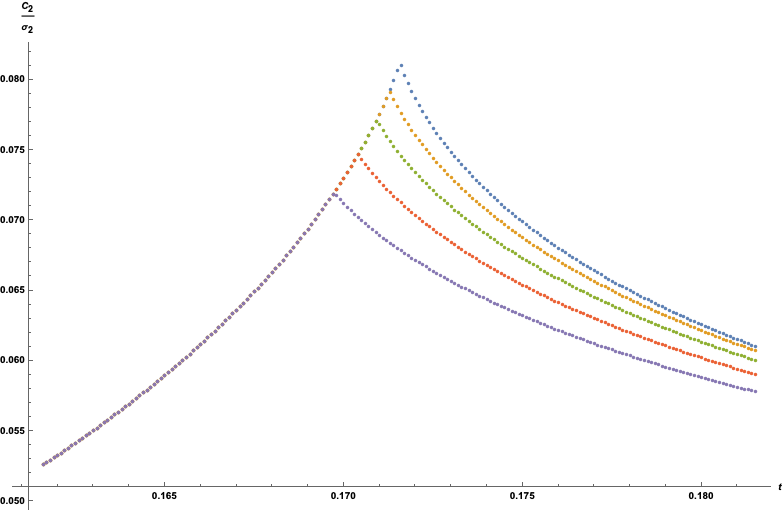}
  \end{subfigure}
  \caption{$C_1$ (left) and $C_2$ (right) vs t for different values of fixed $\overline{\mu}$. From top to bottom, $\overline{\mu} = \sqrt{2} - 1, \sqrt{2} - 1.02, \sqrt{2} - 1.03, \sqrt{2} - 1.04, \sqrt{2} - 1.05$. These are for d=2 and $\sigma_2 =$ $100 \times 100$.}
  \label{fig2}
\end{figure*}

\begin{figure*}[h]
  \centering
  \begin{subfigure}[t]{0.4\textwidth}
  \raggedleft
  \includegraphics[width=\textwidth,height=5cm]{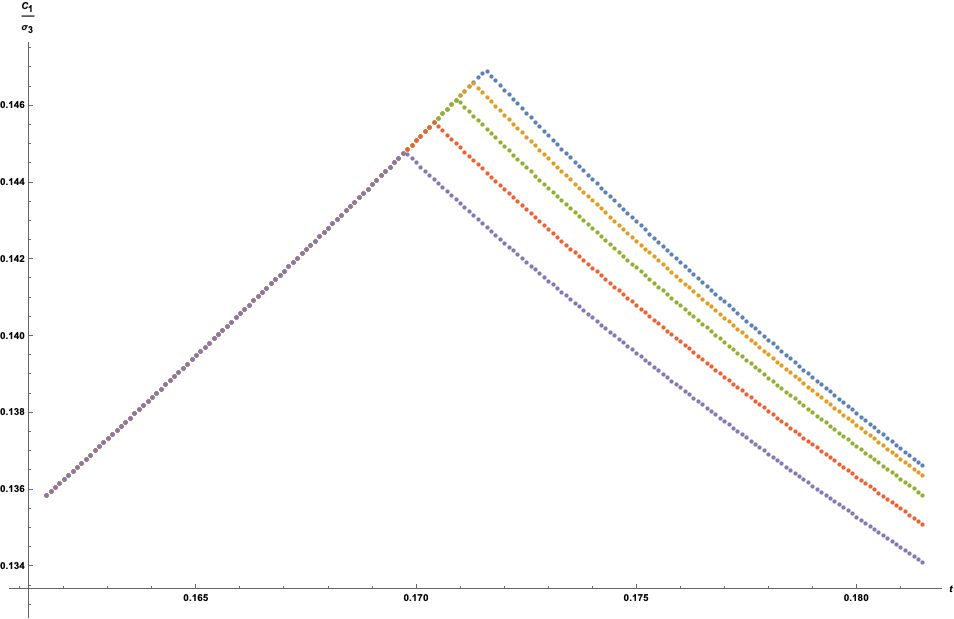}
  \end{subfigure}
  \hfill
  \begin{subfigure}[t]{0.4\textwidth}
  \raggedright
  \includegraphics[width=\textwidth,height=5cm]{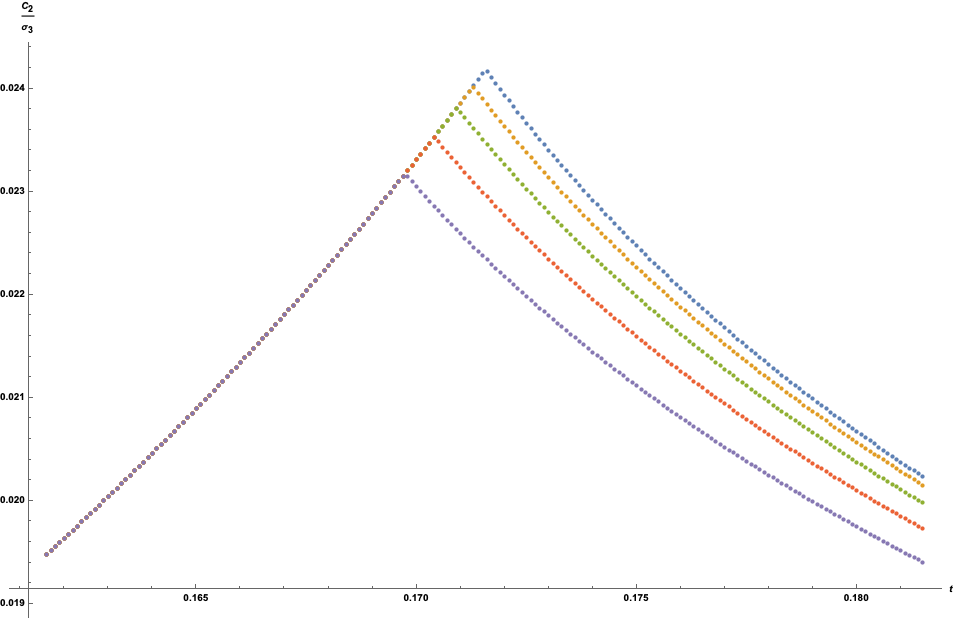}
  \end{subfigure}
  \caption{$\kappa=1,2$ Complexities for fixed $\overline{\mu}$ in the near n=1 lobe region of the phase diagram for d=3. From top to bottom, $\overline{\mu} = \sqrt{2} - 1, \sqrt{2} - 1.02,\sqrt{2} - 1.03, \sqrt{2} - 1.04, \sqrt{2} - 1.05$. Here, $\sigma_3 = 20 \times 20 \times 20$.}
  \label{fig4}
\end{figure*}

Having established that the nature of these excitations is in accordance with those expected from the BH model, we compute the complexity contributions to $C_{\kappa}$ from all of the modes in the lattice system numerically for $\kappa = 1$ and $2$. These complexities are relative to a product state, the mean field ground state. The complexities $C_{\kappa}$ are proportional to the number of sites. So we normalize $C_{\kappa}$ for the ground state by the number of sites and plot $C_\kappa$ vs. $t$ for fixed $\bar{\mu}$ in Figure \ref{fig2} for $d=2$. 
The results indicate that the $O(2)$ critical points (tip of the Mott lobe) have the most complex ground states across the whole $t-\overline{\mu}$ plane. This is also checked for the $n=1$ lobe region in $d=3$ in Figure \ref{fig4}.
This shows that the complexity defined in \eqref{d1} is sensitive not only to the system becoming critical but also to the kind of criticality our bosonic lattice model exhibits. This is not surprising. As we discussed previously, the two kinds of phase transitions differ in the nature of low energy modes and the complexity of the state is sensitive to these modes since they are harder to construct using one and two--site gates.
\begin{figure*}[h]
	\centering
	\begin{subfigure}[t]{0.4\textwidth}
		\raggedleft
		\includegraphics[width=\textwidth,height=5cm]{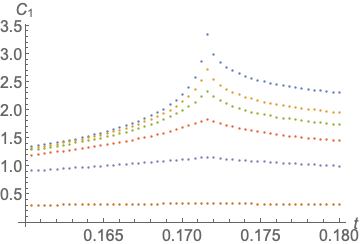}
	\end{subfigure}
	\hfill
	\begin{subfigure}[t]{0.4\textwidth}
		\raggedright
		\includegraphics[width=\textwidth,height=5cm]{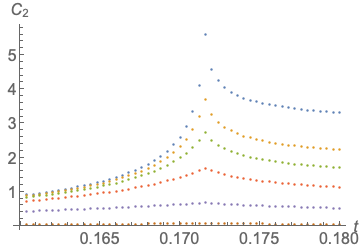}
	\end{subfigure}
	\caption{Complexity contribution from different momentum modes in $d=2$. Each branch is a different value of the momentum. The momenta branches included all have $k_1=0$ and, from the bottom up, $k_2=\pi/2, \pi/5, \pi/10, 3\pi/50, \pi/25, \pi/50$. Each point involves a sum over boson flavors at fixed momentum and $\overline{\mu} = \overline{\mu}_c$. The lattice is $100 \times 100$.}
	\label{fig5}
\end{figure*}

\subsection{Field-theoretic Gaussian prediction}
Now we compute the complexity of the fluctuations near the $O(2)$ critical point. Since we expect the low energy physics to be dominated by the lowest energy modes, the peak in the complexity is determined by them. Although the other modes also have a large contribution to the complexity we expect their contribution to be smooth without a peak at the transition and can be thought as a ``background''.  For the two lowest modes, the mass behaves as $m\sim |t-t_c|^\nu$ where $\nu$ is the usual critical exponent for the correlation length $\xi\sim |t-t_c|^{-\nu}$. In this case, using the Gaussian field theory approximation \footnote{In general the Gaussian approximation to the O(N) model will have a factor of $N/2^{k}$}
\beq
 C_\kappa = \frac{1}{2^{\kappa-1}} \sum_p \left|\ln\left(\frac{\sqrt{p^2+m^2}}{\omega_0}\right)\right|^\kappa = \frac{1}{2^{\kappa-1}}\frac{V\Omega_{d-1}}{(2\pi)^{d}} \int_0^\Lambda p^{d-1} dp \left|\ln\left(\frac{\sqrt{p^2+m^2}}{\omega_0}\right)\right|^\kappa
\label{a65}
\eeq 
where $V$ is the volume of the system and $\Omega_{d-1}= \frac{2\pi^{d/2}}{\Gamma(d/2)}$ is the volume of the unit sphere $S^{d-1}$. A natural value for the momentum cut-off is $\Lambda=\sqrt{\omega_0^2-m^2}$ where the logarithm vanishes. For $\omega_0\gg m$ we can take $\omega_0\sim \frac{1}{a}$ where $a$ is the lattice spacing. In the quantum field theory we take $\omega_0$ as a UV cut-off. We get
\beq
 c_\kappa = \frac{C_\kappa}{V} = \frac{1}{2^{\kappa-1}}\frac{\Omega_{d-1}}{(2\pi)^{d}} \int_0^{\sqrt{\omega_0^2-m^2}} p^{d-1} dp \left|\ln\left(\frac{\sqrt{p^2+m^2}}{\omega_0}\right)\right|^\kappa
\label{a66}
\eeq
Notice that different values of $\kappa$ are related by the simple recursion
\beq
 c_{\kappa-1} = \frac{2\omega_0}{\kappa} \frac{\partial c_\kappa}{\partial\omega_0}, \ \ \ \ c_\kappa = \frac{\kappa}{2} \int_m^{\omega_0} c_{\kappa-1}(\omega'_0)\frac{d\omega'_0}{\omega'_0}
\label{a67}
\eeq
Therefore, the knowledge of the qubit-complexity $C_{QC}$ determines $c_2$ and consequently all the other $c_{\kappa}$ using the recursion relation.
A direct computation of $c_1$ gives
\beq
c_1 = \frac{\Omega_{d-1}}{(2\pi)^dm^2}\frac{(\omega_0^2-m^2)^{1+d/2}}{d(d+2)} {}_2 F_{1}[1,1+d/2,2+d/2,1-\frac{\omega_0^2}{m^2}]
\label{a68}
\eeq
For $d$ even:
\begin{align}
  c_1 =& \frac{1}{2d}\frac{\Omega_{d-1}}{(2\pi)^d}\left\{(-1)^{\frac{d}{2}}m^d\left(-\ln\frac{m^2}{\omega_0^2}+\Psi(1+\frac{d}{2})+\gamma\right) \right.  \label{a69}   \\ 
       &   \hspace{150pt} \left. + \hspace{10pt} \omega_0^d \sum_{n=0}^{\frac{d}{2}-1}\frac{(-1)^n}{\frac{d}{2}-n}\left(\begin{array}{c} d/2\\ n\end{array}\right)\left(\frac{m}{\omega_0}\right)^{2n} \right\} \non  \\
  c_2 =& \frac{1}{2d} \frac{\Omega_{d-1}}{(2\pi)^d} \left\{ m^d\left( (-1)^{\frac{d}{2}}\ln^2\frac{\omega_0}{m}+(-1)^{\frac{d}{2}} (\Psi(1+\frac{d}{2})+\gamma) \ln\frac{\omega_0}{m} \right. \right. \\
       & \left. \left. -\half\sum_{n=0}^{\frac{d}{2}-1} (-1)^n \left(\begin{array}{c} d/2\\ n\end{array}\right) \frac{1}{\left(\frac{d}{2}-n\right)^2} \right) +\frac{\omega_0^d}{2} \sum_{n=0}^{\frac{d}{2}-1} (-1)^n \left(\begin{array}{c} d/2\\ n\end{array}\right) \frac{1}{\left(\frac{d}{2}-n\right)^2}\left(\frac{m}{\omega_0} \right)^{2n}\right\}   \non
\end{align}
and for $d$ odd:
\begin{align}
 c_1 =& \frac{1}{4}\frac{\Omega_{d-1}}{(2\pi)^d}\left\{\omega_0^d\sum_{n=0}^\infty \frac{(-1)^n}{(\frac{d}{2}-n)} \frac{\Gamma(\frac{d}{2})}{n!\Gamma(\frac{d}{2}-n+1)}\left(\frac{m}{\omega_0}\right)^{2n} \right.  \non  \\
 & \hspace{200pt} \left.-(-1)^{\frac{d-1}{2}}\frac{2\pi}{d}m^d\right\} \label{a70} \\
  c_2 =& \frac{1}{4} \frac{\Omega_{d-1}}{(2\pi)^d} \left\{\half \omega_0^d \sum_{n=0}^\infty \frac{(-1)^n}{n!} \frac{\Gamma(\frac{d}{2})}{\Gamma(\frac{d}{2}-n+1)}  \frac{1}{(\frac{d}{2}-n)^2} \left(\frac{m}{\omega_0}\right)^{2n} \right.   \label{a71} \\
 & \left. -\half m^d \sum_{n=0}^\infty \frac{(-1)^n}{n!} \frac{\Gamma(\frac{d}{2})}{\Gamma(\frac{d}{2}-n+1)} \frac{1}{(\frac{d}{2}-n)^2}-(-1)^{\frac{d-1}{2}} \frac{2\pi}{d} m^d\ln\frac{\omega_0}{m} \right\}   \non
\end{align}
Here $\Psi$ is the digamma function and $\gamma$ is the Euler constant. Both expressions are expansions in even powers of $m$ plus a term with $m^d \ln m$ for $d$ even and $m^d$ for $d$ odd. In our case we are interested in $d=2$:
\begin{align}
 c^{d=2}_1 =& \frac{1}{8\pi} \left\{\boxed{m^2\ln m^2} -m^2\ln{\omega_0^2}-m^2+\omega_0^2\right\} \\
 c^{d=2}_2 =& \frac{1}{16\pi} \left\{\boxed{-\frac{m^2}{2}\ln^2 m^2} + m^2\ln m^2 \ln \omega_0^2 -\frac{m^2}{2}\ln^2\omega_0^2  + m^2\ln\frac{m^2}{\omega_0^2}-m^2+\omega_0^2\right\}
\label{a72}
\end{align}
and $d=3$:
\begin{align}
c^{d=3}_1 =& \frac{(\omega_0^2 -m^2)^{-1/2}}{6 \pi^2}\left( \frac{\omega_0^4}{3} -\frac{5}{3}m^2\omega_0^2 + \frac{4m^2}{3} + m^3 \sqrt{\omega_0^2 - m^2} \sin^{-1}{\sqrt{1-\frac{m^2}{\omega_0^2}}} \right) \non \\
 =& \frac{\omega_0^3}{18\pi^2} -\frac{m^2\omega_0}{4\pi^2} +\boxed{\frac{m^3}{12\pi}} +\cO(m^4) \\
c^{d=3}_2 =& \frac{\omega_0^3}{54\pi^2} - \frac{1}{4\pi^2} \omega_0 m^2 - \boxed{\frac{m^3}{24\pi} \ln m^2 } + \frac{m^3}{36\pi}\left(\frac{3}{2} \ln \omega_0^2 -\frac{3}{2}\ln4 + 4\right) + \cO(m^4)
\label{a73}
\end{align}
The quantity $\delta c = c(t_c) - c(t)$ depends on the UV cutoff $\omega_0$ but there is a universal, or cut-off independent, sub-leading piece in the series expression for $\delta c$ that we framed in the equations. If we set $m\sim \sqrt{t-t_c}$ as corresponds to the gaussian model, this universal sub-leading piece has a non-analytic behavior that will dominate higher enough derivatives $d^p \delta c/dt^p$ as $t\rightarrow t_c$. 
This feature is something that one also finds in calculations of complexity that use the holographic proposals namely the CV and CA proposals \cite{25,26,29,30} in geometries that represent RG flows near criticality. We discuss this in a future work \cite{44}.    

\subsection{Summary of numerical results}
The growth in the complexities as we approach criticality is dominated by the part that comes from small momentum. This is shown in Figure \ref{fig5}. The contribution from larger values of the lattice momenta do contribute to the complexity but not to the growth near-criticality. Only the complexity coming from the small momentum modes is sensitive to $t-t_c$. \\
The point in the phase diagram which is at the tip of the lobe is known to have a particle-hole symmetry and both of these excitations become massless at this critical point. The complexities are largest at this point because the contributions from both these excitations grow sharply as we approach this point. At the other generic critical points, only one of the two excitations become massless.

\begin{figure*}[htbp]
  \centering
  \begin{subfigure}[t]{0.4\textwidth}
  \raggedleft
  \includegraphics[width=\textwidth,height=3cm]{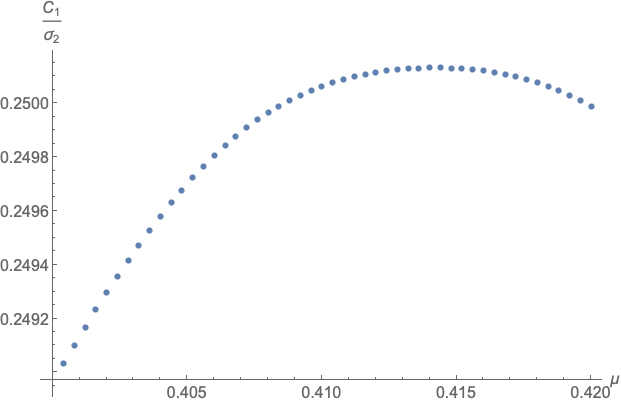}
  \caption{$\kappa=1$ complexity at fixed $t=t_c$.}
  \end{subfigure}
  \hfill
  \begin{subfigure}[t]{0.4\textwidth}
  \raggedright
  \includegraphics[width=\textwidth,height=3cm]{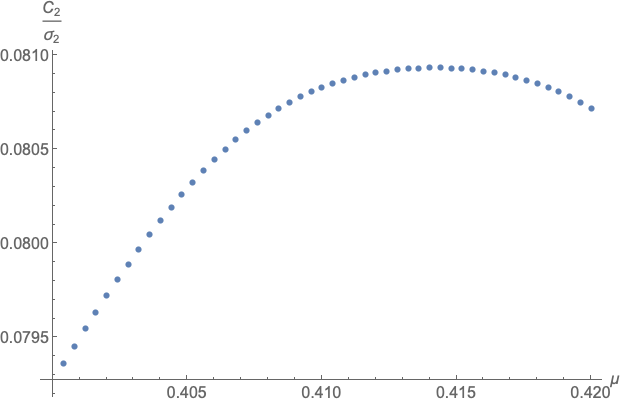}
  \caption{$\kappa=2$ complexity at fixed $t=t_c$.}
  \end{subfigure} 
  \caption{$100 \times 100$ lattice with $d=2$.}
  \label{fig7}
\end{figure*}

\begin{figure*}[htbp]
  \centering
  \begin{subfigure}[t]{0.4\textwidth}
  \raggedleft
  \includegraphics[width=\textwidth,height=4cm]{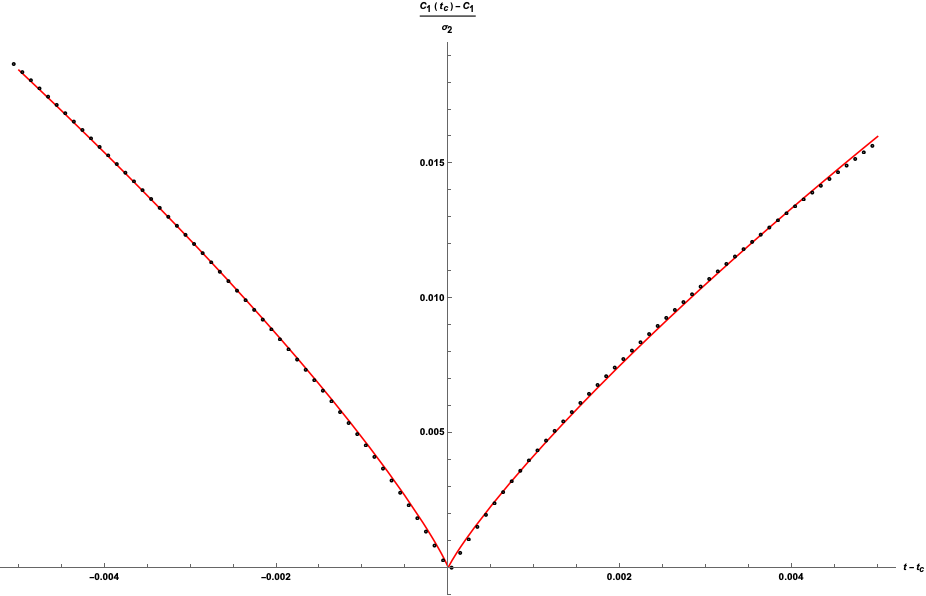}
  \caption{\commentout{The solid line is the function $-\upsilon_{1\pm}\abs{t-t_c} \log{\abs{t-t_c}}$ with
  $\begin{cases} 
        \upsilon_{1-}=0.6968 & t-t_c \leq 0 \\
        \upsilon_{1+}=0.6039  & t-t_c\geq 0 
     \end{cases}$\\
     and $C_1 (t_c)/\sigma_2 = 0.2566$.
    }}
  \end{subfigure}
  \hfill
  \begin{subfigure}[t]{0.4\textwidth}
  \raggedright
  \includegraphics[width=\textwidth,height=4cm]{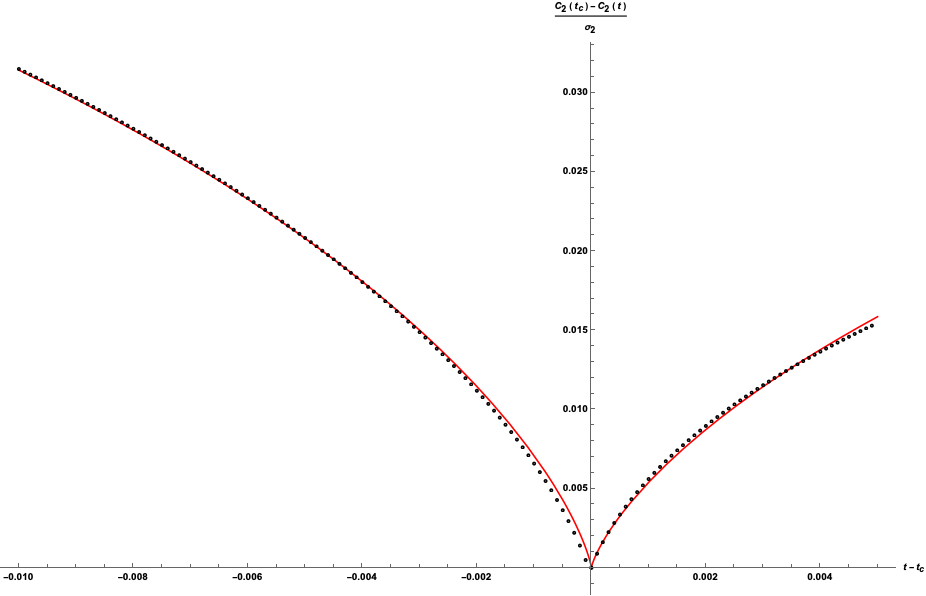}
  \caption{\commentout{The solid line is the function $\upsilon_{2\pm}\abs{t-t_c} (\log{\abs{t-t_c}})^2$ with
  $\begin{cases} 
        \upsilon_{2-}=0.1467 & t-t_c \leq 0 \\
        \upsilon_{2+}=0.1129 & t-t_c\geq 0 
     \end{cases}$\\
     and $C_2(t_c)/\sigma_2 = 0.0861$
  }}
  \end{subfigure} \\
  \begin{subfigure}[t]{0.4\textwidth}
  \raggedright
  \includegraphics[width=\textwidth,height=4cm]{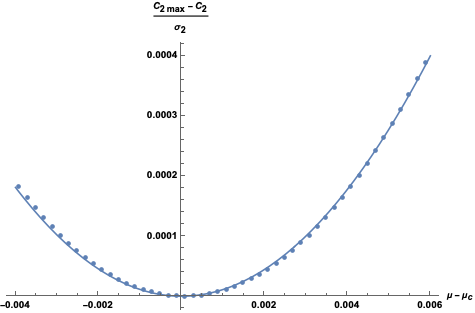}
  \caption{\commentout{The solid line is $d_{2\pm}(\mu-\mu_c)^2$ with
  $\begin{cases} 
        d_{2-}=11.3025 & \mu-\mu_c \leq 0 \\
        d_{2+}=11.0965  & \mu-\mu_c\geq 0 
     \end{cases}$
    }}
  \end{subfigure}
  \hfill
  \begin{subfigure}[t]{0.4\textwidth}
  \raggedright
  \includegraphics[width=\textwidth,height=4cm]{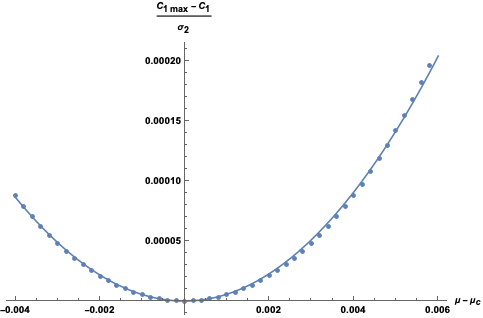}
  \caption{\commentout{The solid line is $d_{1\pm}(\mu-\mu_c)^2$ with
  $\begin{cases} 
        d_{1-}=5.397 & \mu-\mu_c \leq 0 \\
        d_{1+}=5.67  & \mu-\mu_c\geq 0 
     \end{cases}$
    }}
  \end{subfigure}
  \caption{The dotted points are the numerical data. The lattice is $200 \times 200$. The solid lines are are the functions}
  $\begin{array}{ c|c c }
    \hline
    (a) & -\upsilon_{1\pm}\abs{t-t_c} \log{\abs{t-t_c}} &  \begin{array}{c c}
      \upsilon_{1-}=0.6968 & t-t_c \leq 0 \\
      \upsilon_{1+}=0.6039  & t-t_c\geq 0 
   \end{array}
    \\
    \hline
    (b) & \upsilon_{2\pm}\abs{t-t_c} (\log{\abs{t-t_c}})^2 & \begin{array}{c c} 
      \upsilon_{2-}=0.1467 & t-t_c \leq 0 \\
      \upsilon_{2+}=0.1129 & t-t_c\geq 0 
   \end{array}  \\
   \hline
   (c) & d_{2\pm}(\mu-\mu_c)^2 & \begin{array}{c c} 
  d_{2-}=11.3025 & \mu-\mu_c \leq 0 \\
  d_{2+}=11.0965  & \mu-\mu_c\geq 0 
  \end{array} \\
  \hline
  (d) & d_{1\pm}(\mu-\mu_c)^2 &  \begin{array}{c c} 
  d_{1-}=5.397 & \mu-\mu_c \leq 0 \\
  d_{1+}=5.67  & \mu-\mu_c\geq 0 
\end{array}  \\
\hline
\end{array} $
  \label{fig6}
  \end{figure*}

\begin{figure*}[htbp]
    \centering
    \begin{subfigure}[t]{0.4\textwidth}
      \raggedright
      \includegraphics[width=\textwidth,height=4cm]{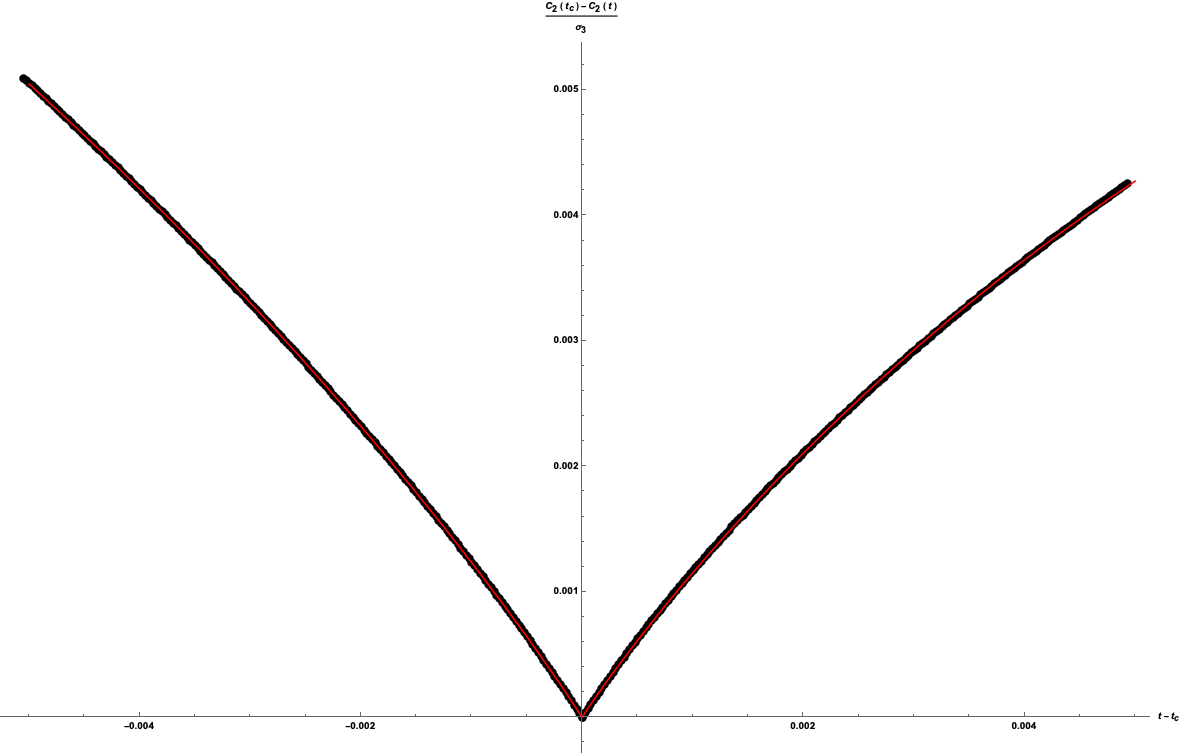}
      \caption{\commentout{The solid line is the function $\upsilon_{2\pm}\abs{t-t_c} (\log{\abs{t-t_c}})^2$ with
      $\begin{cases} 
            \upsilon_{2-}=0.1467 & t-t_c \leq 0 \\
            \upsilon_{2+}=0.1129 & t-t_c\geq 0 
         \end{cases}$\\
         and $C_2(t_c)/\sigma_2 = 0.0861$
      }}
      \end{subfigure}
      \begin{subfigure}[t]{0.4\textwidth}
      \raggedleft
      \includegraphics[width=\textwidth,height=4cm]{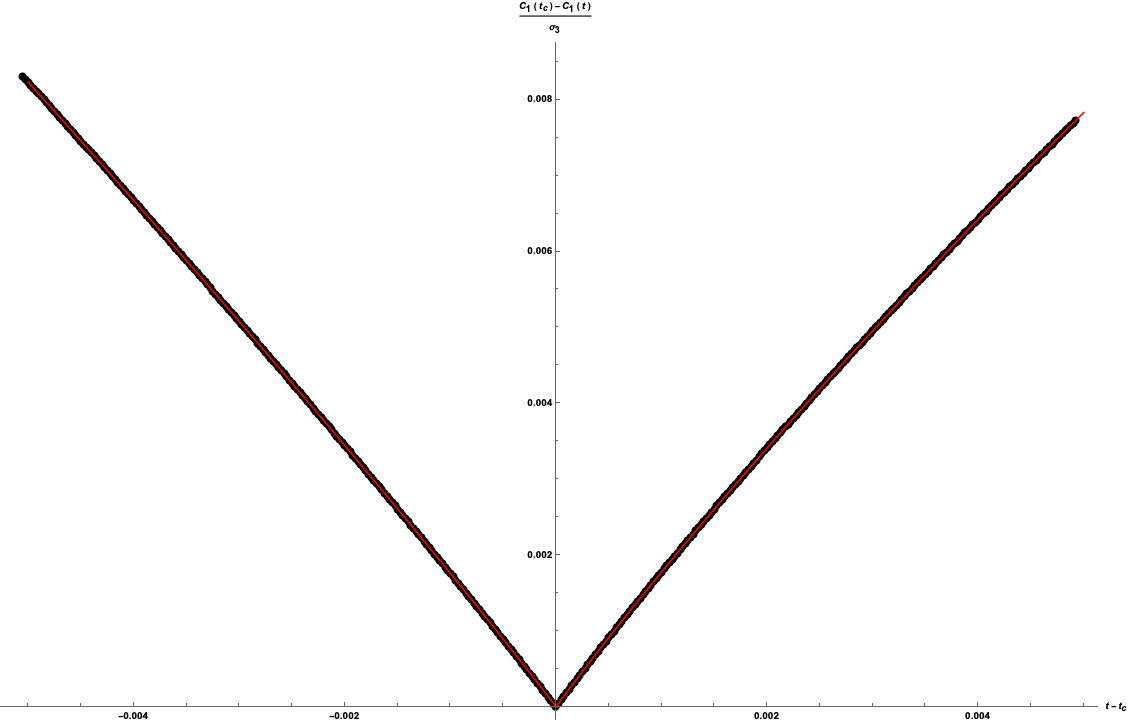}
      \caption{\commentout{The solid line is the function $a_{\pm} \abs{t-t_c} + b_{\pm} \abs{t-t_c}^{3/2}$ with
      $\begin{cases} 
            a_{-}=1.8468, b_{-}=-2.83 & t-t_c \leq 0 \\
            a_{+}=1.944, b_{+}=-5.3425  & t-t_c\geq 0 
         \end{cases}$\\
         and $C_1(t_c)/\sigma_3 = 0.1812$.
        }}
      \end{subfigure}
      \caption{The dotted points are the numerical data. The lattice is $100 \times 100 \times 100$. The solid lines are are the functions}
      $\begin{array}{ c|c c }
        \hline
       (a) & a_{\pm} \abs{t-t_c} + b_{\pm} \abs{t-t_c}^{3/2} & \begin{array}{c c c} 
        a_{-}=1.418 ,& b_{-}=-5.776 & t-t_c \leq 0 \\
        a_{+}=1.393,& b_{+}=-7.618  & t-t_c\geq 0 
      \end{array} \\
      \hline
      (b) & a_{\pm} \abs{t-t_c} + b_{\pm} \abs{t-t_c}^{3/2} & \begin{array}{c c c} 
        a_{-}=1.418 ,& b_{-}=-5.776 & t-t_c \leq 0 \\
        a_{+}=1.393 ,& b_{+}=-7.618  & t-t_c\geq 0 
     \end{array}  \\
     \hline
    \end{array} $
      \label{fig11}
    \end{figure*}

One can also approach the multicritical point at the tip of the lobe by fixing the hopping $t$ and tuning $\mu$. This approach is tangent to the boundary of the lobe at the critical point. One instance of the behaviour of the complexity in this case is shown in Figure \ref{fig7}. Again, the point of the maximal complexity along this trajectory agrees with the critical point. However, the behaviour near the critical point is very different from Figure \ref{fig2}. It is known that several properties of the multicritical point (for ex. the critical exponents) depend on the direction from which it is approached \cite{19}. The complexities are another such property. In Figure \ref{fig6} and \ref{fig11}, we show the behavior found near the $O(2)$ critical point using numerics along with the expected field theory behaviors with classical exponents for $d=2$ and $d=3$ resepectively. We find that the two scaling behaviors match and find the coefficients needed for the matching to occur. 

\begin{figure*}[h]
  \centering
  \includegraphics[width=\textwidth,height=6cm]{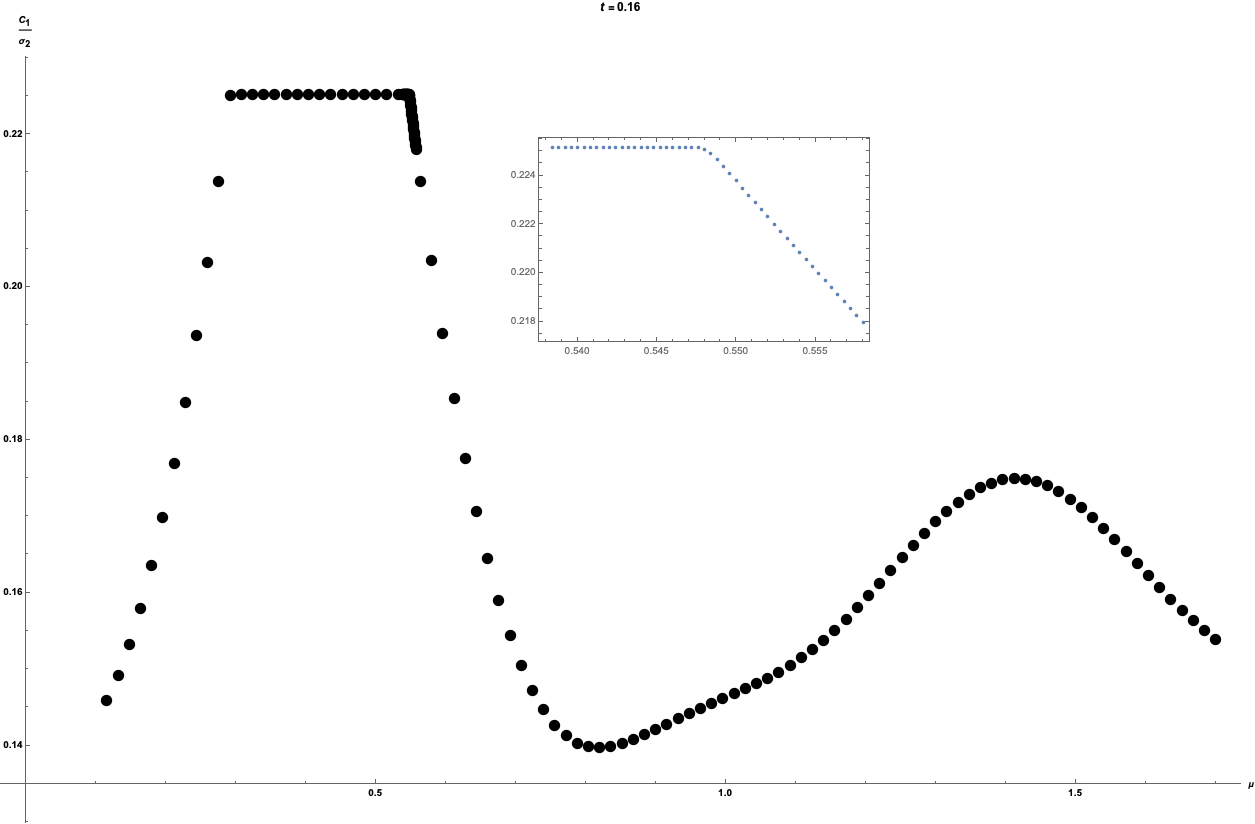}
  \caption{$\kappa=1$ complexity at fixed $t=0.16$. $100 \times 100$ lattice with $d=2$. The inset shows the complexity near phase transition from the Mott insulator phase (left) to the superfluid at $\mu=0.548$. The decrease in complexity signifies the onset of superfluidity.}
  \label{fig8}
\end{figure*}

\begin{figure*}[h]
  \centering
  \begin{subfigure}[t]{0.4\textwidth}
  \raggedleft
  \includegraphics[width=\textwidth,height=5cm]{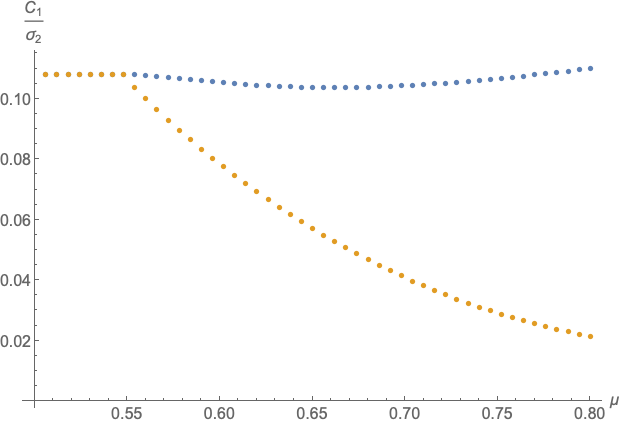}
  \caption{$C_1$ contributions at the generic transition at $t=0.16$. The blue line is the excitation that becomes massless in the SF while the yellow line is the gapped one.}
  \end{subfigure}
  \hfill
  \begin{subfigure}[t]{0.4\textwidth}
  \raggedright
  \includegraphics[width=\textwidth,height=5cm]{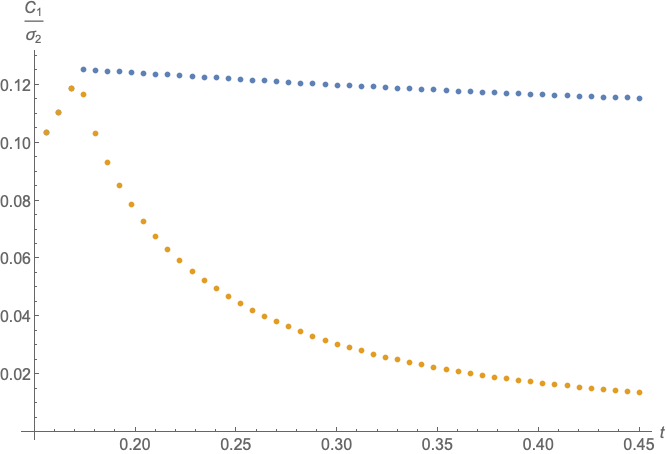}
  \caption{$C_1$ contributions at the O(2) transition at $\mu=0.4142$. Both of the modes become massless at the critical point. Inside the SF, the gap reopens for the yellow one but not for the blue one.}
  \end{subfigure} 
  \caption{$100 \times 100$ lattice with $d=2$ and $f=4$. $\kappa=1$ complexity for the two lowest excitations at fixed $t$ in $(a)$ and fixed $\mu$ in $(b)$.}
  \label{fig9}
\end{figure*}

An interesting aspect that arises in Figures \ref{fig3}, \ref{fig2}, \ref{fig4} is that inside the Mott lobes, the complexities are independent of $\mu$. This can be explicitly checked  by looking at complexity at fixed t inside the lobes and varying $\mu$. This is shown in Figure \ref{fig8}. We find that complexity is independent of $\mu$ inside the lobes but only depends on t. To what extent this holds beyond the approximation we made in this paper remains to be seen. \\
 Figure \ref{fig8} also has an unexpected feature - in the generic MI-SF transition, the complexity decreases on the SF side. To better understand this, we can look at the complexity contributions coming from the different boson flavors. In the MI phase, we find that the two lowest boson flavors which correspond to the particle and hole excitations of the Bose-Hubbard model contribute significantly while the complexity of the more energetic flavors is smaller by a few orders of magnitude and can be neglected. Thus, we can study the behaviour of the complexity of these two flavors as we approach the phase transition in both the $O(2)$ critical and the generic case from the MI side. In the generic case, even though one of the two flavors becomes gapless leading to a larger complexity, the mass of the other excitation grows with the deformation and its complexity decreases. In fact this decrease overwhelms the increase leading to an overall decrease which can be seen in Figure \ref{fig8}. The behavior of the complexity of the two flavors near the phase transition is shown in Figure \ref{fig9}(a). In the $O(2)$ criticality case, as we move towards the critical point from the MI side the complexities see a sharp increase as both flavors become gapless. On the other side towards the SF, there is a decrease but not as sharp since the decrease comes only from one of the two flavors, the one whose gap opens up on the SF side while the complexity of the gapless mode does not change near the phase transition on the superfluid side. This is shown in Figure \ref{fig9}(b).

\section{Complexity-Volume proposal in Holography}\label{sec7}
\begin{figure}[h]
	\centering
	\includegraphics[height=8cm]{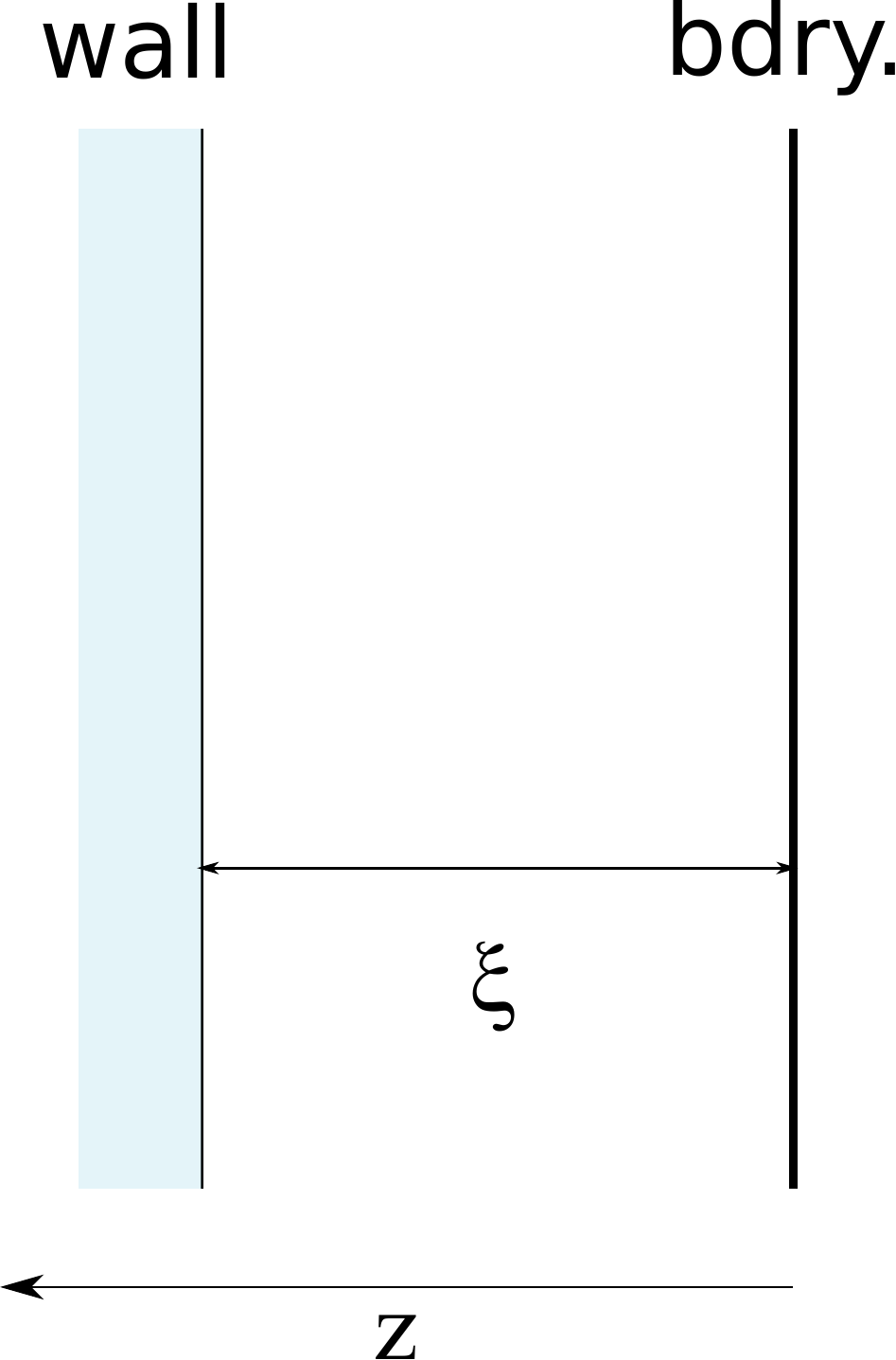}
	\caption{Schematic AdS/CFT model for a near criticality. At criticality the space is AdS and the wall provides a mass scale.} 
	\label{fig10}
\end{figure}

The geometrization of RG flows in the context of the AdS/CFT correspondence is well known \cite{38}-\cite{41}. There exist several constructions of  boundary field theories in $(d+1)$ spacetime dimensions that flow from a UV fixed point to gapped theories in the IR or to other IR fixed points \cite{42},\cite{43}. 
Critical theories deformed by relevant scalar operators $\cO$ with $\Delta < d+1$ are usually studied in holography by introducing a bulk scalar field $\Phi$ to the gravitational theory dual to the boundary operator $\cO$. The near boundary $(z \rightarrow 0)$ behaviour of $\Phi$ is given by $\Phi \sim \Phi_{(s)}z^{d+1-\Delta}$ corresponding to a source $\Phi_{(s)} \sim t-t_c$ added to the action of the boundary theory. The field $\Phi$ starts out as a fluctuation near the boundary that goes to $0$ as $z \rightarrow 0$ since such constructions have an asymptotic AdS spacetime in $(d+2)$-dimensions. The scalar grows as we proceed into the bulk due to the relevant nature of the deformation. In fact, it is no longer a small fluctuation when $z \sim |t-t_c|^{-\frac{1}{d+1-\Delta}} = |t-t_c|^{-\nu}$ or equivalently $z \sim \xi$. In this range of the bulk coordinate, the bulk field has grown significantly so that it can back-react strongly and change the metric. For gapped theories, the change in the metric is of such a form that the spacetime ends with some internal manifold smoothly capping off in a higher dimensional supergravity model in which the lower dimensional theory is embedded. For  simplicity, we can model this phenomena by ending our ($d+2$)-dimensional spacetime by a wall as shown in Figure \ref{fig10}.
In this section, we consider the qualitative behaviour of the holographic complexity in a simple situation and extract the kind of scaling already seen in Fig. \ref{fig6}. We leave a more detailed analysis of holographic complexity using both the CV and CA prescriptions for exact solutions of known holographic fixed points and closely related toy models to a future study \cite{44}.
For a deformation away from the critical point that in the IR flows to a gapped theory, the basic setup is an AdS space with a wall at some value of the radial coordinate. Consider the AdS metric in Poincare coordinates,
\begin{align}
ds^2 = \frac{L^2}{z^2} (\eta_{\mu \nu}dx^{\mu}dx^{\nu} + dz^2)
\label{a84}
\end{align}
Here, $\eta_{\mu \nu}$ is the standard Minkowski metric in $(d+1)$ dimensions and $x$ are the boundary spacetime coordinates on which the boundary field theory is defined. Here $z$ is a radial coordinate that extends into the bulk and runs from $z=0$ to $z=\xi$. The spacetime ends at $z=\xi$ instead of going all the way to $z=\infty$. The fact that there are no degrees of freedom in the field theory with energy less than the gap corresponds to the spacetime ending at this particular value of the $z$ coordinate. This is consistent with the identification of IR phenomenon in the field theory with large $z$ phenomenon in the bulk \cite{44b}. 
The CV conjecture \cite{25} allows for a geometric method to calculate the complexity in holographic theories. The conjecture is that the complexity of a field theory state on a fixed time slice can be found by computing the bulk codimension-1 volume of the maximal Cauchy slice that is anchored on the same fixed time boundary slice.
\begin{align}
C_V =  \frac{V_{\Sigma}}{G_N l}
\label{a85}
\end{align}
Here $\Sigma$ is the maximal Cauchy slice satisfying the correct boundary condition, $l$ is some length scale associated with the geometry usually taken to be the AdS length scale L and $G_N$ is the gravitational constant in $(d+2)$-dimensions. We also make the choice $l=L$.
The above spacetime has a time-translation symmetry and we can consider slices anchored at $t=0$ on the boundary without loss of generality. Moreover, the maximal slices are just fixed time slices. By construction, the maximal complexity occurs for the geometry dual to the fixed point when $\xi \rightarrow \infty$ and is proportional to the spatial field theory volume $\sigma_{d}$.  The complexity is UV divergent because the maximal slice goes all the way to the boundary where the AdS metric diverges and so we regulate the geometry at $z=\epsilon$. A simple computation yields, 
\begin{align}
\delta C_V(\xi) = C(\xi \rightarrow \infty) - C(\xi) = \frac{\sigma_{d}L^{d}}{dG_N^{(d+2)}} \frac{1}{\xi^{d}}
\label{a86}
\end{align}
The first factor in the holographic field theory translates into a measure of the number of degrees of freedom in the boundary theory. For example in the case of $AdS_5$, the ratio $L^{d}/G^{(5)}_N$ is proportional to $N^2$ where $N$ is the rank of the gauge group of the dual theory and in the case of $AdS_3$, $L/G^{(3)}_N$ is proportional to the central charge of the holographic boundary conformal field theory. Using the definition of the critical exponent $\nu$, we have 
$\xi \sim |t-t_c|^{-\nu}$. Since the dynamical exponent in this case is one due to Lorentz symmetry in the boundary theory, we have $m \sim 1/\xi$. So  $\delta C_V$ is proportional to $m^{d}$ or $|t-t_c|^{\nu d}$ in this simple holographic case in agreement with the scaling behavior expected from general arguments. 

\section{Conclusions}\label{sec8}
 
 In this work we studied numerically and analytically the complexity of the ground state of the Bose-Hubbard model in two and three dimensions in the Mott and superfluid phases as well as near the critical point. First we mapped the model to a system of qubits for which we apply the standard definition of complexity \eqref{a43} obtaining a general formula \eqref{a44} in terms of a complexity Hamiltonian \eqref{twoqudit} that depends on the target and reference states. Numerically we compute the ground state by using a mean field approximation plus fluctuations. The mean field ground state is taken as the reference state and the target state is a condensate of fluctuations that is related to the reference state by a Bogoliubov transformation in which case the complexity Hamiltonian \eqref{a42} is independent of the auxiliary evolution parameter $\tau$ leading to the simple expression \eqref{a45}. In this way we numerically evaluate the complexity and find that it has a maximum at the $O(2)$ critical point which is one of our main results. However, very close to the critical point this approximation is not valid. On the other hand, in that region the only scale is given by the correlation length and general scaling arguments allow us to find the dependence of the complexity with the coupling in terms of the standard $\nu$ critical exponent. We verify that our numerical results agree with a classical exponent $\nu=\half$. That the complexity has a maximum at the critical point is related to the difficulty in creating long range correlations using one and two--site gates. We verify this idea by doing an analytical calculation of the complexity for a free field theory of two scalars with masses $m_{1,2}$. 
 When one or both masses tend to zero $m_{1,2}\rightarrow 0$ the complexity has a maximum and, if we use the classical exponent $m\sim\sqrt{|t-t_c|}$ we reproduce the numerical results showing that the maximum of the complexity is due to the massless modes. The full value of the complexity is not necessarily dominated by these modes but the contribution from the other modes across the transition does not have a peak. The complexity depends on a UV cut-off, but whether we use the classical or the exact exponent $m\sim |t-t_c|^\nu$ we find that the complexity near the critical point $C(t\rightarrow t_c)$  has a universal \ie\ cut-off independent, non-analytic behavior that would lead to the divergence of large enough derivatives $\frac{\partial^k C(t)}{\partial t^k}$ as $t\rightarrow t_c$. This is the other main result of the paper. To provide further verification we computed the complexity analytically for systems that have a gravity dual using a simplified gravity model and recent progress in the AdS/CFT correspondence on holographic complexity. The results we obtain agree with our general picture based on the Hubbard model and free field theory. However it is worth emphasizing that our numerical calculation uses the standard definition of circuit complexity in terms of qubits whereas the AdS/CFT calculation uses a different definition conjectured in \cite{25}--\cite{30} to agree with the standard one. 
 
 To improve the numerical calculation beyond our approximation, one should look at the full qubit model with the target state being the ground state of the full Hamiltonian in \eqref{bhqubith}. As mentioned, the complexity Hamiltonian \eqref{twoqudit} is universal and can be used to create any state. Therefore the complexity definition \eqref{a44} is still valid but the parameters $\Theta_{IJ,\alpha\beta\gamma}$ generically depend on $t$ and \eqref{a45} is no longer true. However as long as the approximate ground state for which we want to compute the complexity and the reference state are related by a Bogoliubov transformation the result \eqref{a49} follows. For example, this can allow to study systems with spatial inhomogeneities, in particular it will be interesting to consider random potentials which can lead to a glass phase and will require averaging the complexity over a random distribution. 
 
 Some recent studies have focused on the scaling of complexity and other information theoretic measures like the entanglement entropy in quenches across critical points \cite{45}-\cite{47}. This work opens up the possibility of numerically studying the scaling of the complexity for the same quench protocols in the Bose-Hubbard model. Regarding the holographic computation of complexity near a critical point we could go further than the preliminary steps in Section \ref{sec7} where a simplified model of the gravity dual was used. Although the lattice model considered in this paper does not have a known gravity dual a qualitative comparison can still be made with those near critical quantum field theories for which holographic calculations are valid.  

\section{Acknowledgements}\label{sec9}
 We are very grateful to Chen-Lung Hung, Sergei Khlebnikov, Nima Lashkari and Rob Myers for comments and discussions.  In addition, we are very grateful to the DOE that supported in part this work through grants DE-SC0007884, DE-SC0019202 and the DOE QuantISED program of the theory consortium ``Intersections of QIS and Theoretical Particle Physics'' at Fermilab, as well as to the Keck Foundation that also provided partial support for this work.


\begin{thebibliography}{46}
 \bibitem{1}
 M. A. Nielsen, \textit{A geometric approach to quantum circuit lower bounds}
arXiv:quant-ph/0502070 [quant-ph]. \\

\bibitem{2}
M. A. Nielsen, M. R. Dowling, M. Gu, and A. M. Doherty, \textit{Quantum Computation as
Geometry} Science 311 (2006) 1133-1135, arXiv:quant-ph/0603161 [quant-ph]. \\

\bibitem{3}
M. A. Nielsen and M. R. Dowling, \textit{The geometry of quantum computation}
arXiv:quant-ph/0701004 [quant-ph]

\bibitem{4}
 R. Jefferson and R. C. Myers, \textit{Circuit complexity in quantum field theory}, JHEP
1710, 107 (2017) doi:10.1007/JHEP10(2017)107 [arXiv:1707.08570 [hep-th]]. \\

\bibitem{5}
S. Chapman, M. P. Heller, H. Marrochio and F. Pastawski, \textit{Toward a Definition of
Complexity for Quantum Field Theory States} , Phys. Rev. Lett. 120, no. 12, 121602
(2018) doi:10.1103/PhysRevLett.120.121602 [arXiv:1707.08582 [hep-th]]. \\

\bibitem{6}
R. Khan, C. Krishnan and S. Sharma, \textit{Circuit Complexity in Fermionic Field
Theory} ,[arXiv:1801.07620 [hep-th]].  \\

\bibitem{7}
L. Hackl and R. C. Myers,  \textit{Circuit complexity for free fermions}, JHEP 1807, 139
(2018) doi:10.1007/JHEP07(2018)139 [arXiv:1803.10638 [hep-th]].   \\

\bibitem{8}
M. Guo, J. Hernandez, R. C. Myers and S. M. Ruan,  \textit{Circuit Complexity for Coherent
States} , JHEP 1810, 011 (2018) doi:10.1007/JHEP10(2018)011 [arXiv:1807.07677
[hep-th]]\\

\bibitem{9}
S. Chapman, J. Eisert, L. Hackl, M. P. Heller, R. Jefferson, H. Marrochio and
R. C. Myers,  \textit{Complexity and entanglement for thermofield double states} ,
[arXiv:1810.05151 [hep-th]].  \\

\bibitem{10}
A. Bhattacharyya, A. Shekar and A. Sinha,  \textit{Circuit complexity in interacting QFTs
and RG flows}, JHEP 1810, 140 (2018) doi:10.1007/JHEP10(2018)140
[arXiv:1808.03105 [hep-th]]  \\

\bibitem{11}
J. Jiang and X. Liu, \textit{Circuit Complexity for Fermionic Thermofield Double states} ,
[arXiv:1812.00193 [hep-th]]


\bibitem{12}
M.Greiner, O.Mandel, T.Esslinger, T.W.H{\"a}nsch, I.Bloch, \textit{Quantum phase transition from a superfluid to a Mott insulator in a gas of ultracold atoms}. Nature 415, 39 (2002)\\

\bibitem{13}
T.St{\"o}ferle, H.Moritz, C.Schori, M.K{\"o}hl, T.Esslinger, \textit{Transition from a Strongly Interacting 1D Superfluid to a Mott Insulator} Phys.Rev.Lett. 92, 130403 (2004)  \\

\bibitem{14}
C.Schori, T.St{\"o}ferle, H.Moritz, M.K{\"o}hl, T.Esslinger, \textit{Excitations of a Superfluid in a Three-Dimensional Optical Lattice} Phys.Rev.Lett. 93, 240402 (2004)  \\

\bibitem{15}
K. Xu, Y. Liu, J.R. Abo-Shaeer, T. Mukaiyama, J.K. Chin, D.E. Miller, W. Ketterle, K.M. Jones, E. Tiesinga, \textit{Sodium Bose-Einstein condensates in an optical lattice} Phys. Rev. A 72, 043604 (2005)  \\

\bibitem{16}
F.Gerbier, S.F{\"o}lling, A.Widera, O.Mandel, I.Bloch, \textit{Probing Number Squeezing of Ultracold Atoms across the Superfluid-Mott Insulator Transition} Phys.Rev.Lett. 96, 090401 (2006)


\bibitem{17}
L. D. Landau and E. M. Lifshitz, \textit{Statistical Mechanics}, Part 2, Volume 9 ,
Sec. 25

\bibitem{18}
D. Jaksch, C. Bruder, J. I. Cirac, C. W. Gardiner, and P. Zoller \textit{Cold Bosonic Atoms in Optical Lattices} Phys.Rev.Lett. 81, (1998) doi :	10.1103/PhysRevLett.81.3108  \\

\bibitem{19}
M. P. A. Fisher et al., Phys. Rev. B 40, 546 (1989)

\bibitem{20}
Sachdev, S. (2011). \textit{Quantum Phase Transitions} (2nd ed.), Cambridge: Cambridge University Press. doi:10.1017/CBO9780511973765

\bibitem{21}
Matthias Vojta, \textit{Quantum Phase Transitions}, Reports on Progress in Physics {\bf 66}, 2069 (2003), doi:10.1088/0034-4885/66/12/r01.

\bibitem{22}
P. Caputa, N. Kundu, M. Miyaji, T. Takayanagi and K. Watanabe, \textit{Anti-de Sitter space from optimization of path integrals in conformal field theories}, Phys. Rev. Lett. 119 (2017) 071602 [arXiv:1703.00456]

\bibitem{23}
A. Bhattacharyya, P. Caputa, S.R. Das, N. Kundu, M. Miyaji and T. Takayanagi, \textit{Path-integral complexity for perturbed CFTs}, JHEP 07 (2018) 086 [arXiv:1804.01999]

\bibitem{24}
P. Caputa, N. Kundu, M. Miyaji, T. Takayanagi and K. Watanabe, \textit{Liouville action as path-integral complexity: from continuous tensor networks to AdS/CFT}, JHEP 11 (2017) 097 [arXiv:1706.07056]

\bibitem{25}
L. Susskind, \textit{Entanglement is not enough}, Fortsch. Phys. 64 (2016) 49 [arXiv:1411.0690]

\bibitem{26}
L. Susskind, \textit{Computational complexity and black hole horizons}, Fortsch. Phys. 64 (2016) 24 [Addendum ibid. 64 (2016) 44] [arXiv:1403.5695] [arXiv:1402.5674]

\bibitem{27}
D. Stanford and L. Susskind, \textit{Complexity and shock wave geometries}, Phys. Rev. D 90 (2014) 126007 [arXiv:1406.2678]

\bibitem{28}
L. Susskind and Y. Zhao, \textit{Switchbacks and the bridge to nowhere}, [arXiv:1408.2823]

\bibitem{29}
A.R. Brown, D.A. Roberts, L. Susskind, B. Swingle and Y. Zhao, \textit{Holographic complexity equals bulk action?}, Phys. Rev. Lett. 116 (2016) 191301 [arXiv:1509.07876]

\bibitem{30}
A.R. Brown, D.A. Roberts, L. Susskind, B. Swingle and Y. Zhao, \textit{Complexity, action and black holes}, Phys. Rev. D 93 (2016) 086006 [arXiv:1512.04993]

\bibitem{31}
D. Carmi, R. Myers and P.Rath, \textit{Comments on Holographic Complexity}, JHEP 1703 (2017) 118 [arXiv:1612.00433]

\bibitem{32}
Rolando Somma, Gerardo Ortiz, Emanuel Knill, and James Gubernatis, \textit{Quantum Simulations of Physics Problems}, arXiv:quant-ph/0304063

\bibitem{33}
Focus issue: Quantum phase transitions, Nature Phys. 4, 167-204 (2008)

\bibitem{34}
D. van Oosten, P. van der Straten and H. T. C. Stoof , \textit{Quantum phases in an optical lattice}, Phys.Rev.A63:053601, 2001, [arXiv:cond-mat/0011108]

\bibitem{35}
D. Podolsky, A. Auerbach, and D. P. Arovas, \textit{Visibility of the amplitude (Higgs) mode in condensed matter}, Phys. Rev. B 84, 174522 (2011), [arXiv:1108.5207]

\bibitem{36}
W. Zwerger, \textit{Anomalous Fluctuations in Phases with a Broken Continuous Symmetry}, Physical Review Letters 92, 027203 (2004), [arXiv:cond-mat/0304153]

\bibitem{37}
Sudip Chakravarty, Bertrand I. Halperin and David R. Nelson, \textit{Low-Temperature Behavior of Two-Dimensional Quantum Antiferromagnets}, Physical Review Letters 60,  1057 (1988)

\bibitem{38}
J. M. Maldacena, \textit{The Large N Limit of Superconformal Field Theories and Supergravity}, Adv.Theor.Math.Phys.2:231-252 (1998), [arXiv:hep-th/9711200]

\bibitem{39}
J. Mcgreevy, \textit{Holographic duality with a view toward many-body physics}, Adv.High Energy Phys.2010:723105 (2010), [arXiv:0909.0518]

\bibitem{40}
K. Skenderis, \textit{Lecture Notes on Holographic Renormalization}, Class.Quant.Grav.19:5849-5876 (2002), [arXiv:hep-th/0209067]

\bibitem{41}
Massimo Bianchi, Daniel Z. Freedman
and Kostas Skenderis, \textit{How to go with an RG Flow}, JHEP 0108:041 (2001), [arXiv:hep-th/0105276]

\bibitem{42}
D.Z. Freedman, S.S. Gubser, K. Pilch, and N.P. Warner, \textit{Continuous distributions of D3-branes
and gauged supergravity}, JHEP 0007:038 (2000), [arXiv:hep-th/9906194]

\bibitem{43}
L. Girardello, M. Petrini, M. Porratic and A. Zaffaronia, \textit{The Supergravity Dual of N = 1 Super Yang-Mills Theory
}, Nucl.Phys. B569 (2000) 451-469, 	[arXiv:hep-th/9909047]

\bibitem{44}
M. Kruczenski, U. Sood, \textit{In preparation.}

\bibitem{44b}
A.~W.~Peet and J.~Polchinski,
\textit{UV / IR relations in AdS dynamics}, 
Phys. Rev. D \textbf{59}, 065011 (1999),
[arXiv:hep-th/9809022 [hep-th]].

\bibitem{45}
Hugo A. Camargo, Pawel Caputa, Diptarka Das, Michal P. Heller, and Ro Jefferson, \textit{Complexity as a novel probe of quantum quenches: universal scalings and purifications}, Phys. Rev. Lett. 122, 081601 (2019), [arXiv:1807.07075]

\bibitem{46}
Sinong Liu, \textit{Complexity and scaling in quantum quench in
1 + 1 dimensional fermionic field theories}, J. High Energ. Phys. 2019, 104 (2019), [arXiv:1902.02945]

\bibitem{47}
Tibra Ali, Arpan Bhattacharyya, S. Shajidul Haque, Eugene H. Kim, Nathan Moynihan, \textit{Post-Quench Evolution of Complexity and Entanglement in a Topological System}, Phys. Lett. B, 811 (2020),135919, [arXiv:1811.05985 [hep-th]]

\bibitem{48}
Ashok Muthukrishnan and C. R. Stroud, Jr, \textit{Multivalued logic gates for quantum computation}, Phys. Rev. A 62, 052309 (2000)

\bibitem{49}
Arpan Bhattacharyya, Pratik Nandy, Aninda Sinha, \textit{Renormalized Circuit Complexity}, Phys. Rev. Lett. 124, 101602 (2020), 	arXiv:1907.08223 [hep-th]

\end{thebibliography}
\end{document}